\documentclass[]{spie}  

\usepackage[dvips]{graphics}
\usepackage[dvips]{graphicx}
\usepackage{amsmath,epsfig,epsf,psfrag,amssymb,amsfonts,latexsym}
\usepackage{verbatim}
\usepackage[mathscr]{eucal}

  \ifx\LabelFigloaded\MYundefined\relax
  \else
   \fi

  \def\LabelFigloaded{\relax}

  \chardef\LabelFigCatAt\the\catcode`\@
  \catcode`\@=11

 \let\LabelFigwlog@ld\wlog
 \def\wlog#1{\relax}

 \ifx\\\MYundefined@
    \let\\\relax
 \fi

  \def\ms@g{\immediate\write16}

 \def\N@wif{\csname newif\endcsname }
 \def\Temp@ {\N@wif\ifIN@}
 \ifx\INN@\MYundefined@
    \else \let\Temp@\relax
 \fi
 \Temp@

  \def\IN@{\expandafter\INN@\expandafter}
  \long\def\INN@0#1@#2@{\long\def\NI@##1#1##2##3\ENDNI@
    {\ifx\m@rker##2\IN@false\else\IN@true\fi}%
     \expandafter\NI@#2@@#1\m@rker\ENDNI@}
  \def\m@rker{\m@@rker}
 
  \newtoks\Initialtoks@  \newtoks\Terminaltoks@
  \def\SPLIT@{\expandafter\SPLITT@\expandafter}
  \def\SPLITT@0#1@#2@{\def\TTILPS@##1#1##2@{%
     \Initialtoks@{##1}\Terminaltoks@{##2}}\expandafter\TTILPS@#2@}

 \def\Shifted@@#1#2#3{\setbox0=\hbox{#3}%
   \raise -\dp0\vbox {\kern-#2%
       \hbox {\kern#1\unhbox0\kern-#1}%
           \kern#2}}

 \newcount\gridcount
 \newbox\auxGridbox@ \newbox\hGridbox@ \newbox\vGridbox@
 \newbox\Labelbox@ \newbox\auxLabelbox@
 \newbox\Coordinatebox@
 \newtoks\Labeltoks@
 \newdimen\Wdd@ \newdimen\Htt@
 \newdimen\Wddd@ \newdimen\Httt@
 
 \def\Wr@{\immediate\write16}

 \newdimen\GL@wd
 \def\GridLineWidth#1{\GL@wd=#1}

 \def\gobble#1{}
 \def\EdgeErr@{\Wr@{}%
      \Wr@{\string\Edges\space argument
      1, 10, 100 or 1000 please\string!}%
      }

 \newcount\Edgect@

 \def\Sweepup#1\endSweepup{}

 \def\SetEdges@{%
    \edef\Zr@@s{\expandafter\gobble\number\Edgect@\empty}%
        \count255=0\Zr@@s\relax
        \ifnum\count255=\z@\else\EdgeErr@\show\tailtest\fi
        \count255=1\Zr@@s\relax
        \ifnum\count255=\Edgect@\relax\else\EdgeErr@\show\leadtest\fi
    \EdgGl@b\edef\Zr@s{\expandafter\gobble\Zr@@s\empty}
    \ifnum\Edgect@>\@ne\relax\EdgGl@b\let\L@Dc\empty
        \else\EdgGl@b\edef\L@Dc{\string.}\fi
    \ifnum\Edgect@>\@ne\relax
        \EdgGl@b\edef\Edgescale@##1{\divide##1 by \Edgect@}%
        \else\EdgGl@b\edef\Edgescale@##1{}\fi
    }

 \def\Edges#1{\Edgect@=#1\relax
     \let\EdgGl@b\global \SetEdges@}

 \Edges{1}

 \def\hhrule{\hrule height \GL@wd\vskip-.\GL@wd}

 \def\hRule@{%
   \advance\gridcount -2%
   \vfil\hhrule\vfil
   \llap{\smash{\raise -2.5pt
     \hbox{\L@Dc\number\gridcount\Zr@s\kern2pt}}}%
   \hhrule
   }

\def\vvrule{\vrule width \GL@wd \kern-\GL@wd}

 \def\vRule@{\advance\gridcount 2%
   \hfil\vvrule\hfil
   \setbox\auxGridbox@=\vbox to 0pt
      {\vskip \Htt@\vskip 2pt
        \hbox to 0pt{\hss\L@Dc\number\gridcount\Zr@s\hss}\vss}%
      \wd\auxGridbox@=0pt \box\auxGridbox@
   \vvrule
   }

 \def\PlaceGrid@@{\gridcount=10 
  \setbox\hGridbox@=\hbox{%
        \hbox{%
             \hskip-.4pt\vrule
             \vbox to \Htt@{%
               \offinterlineskip\parindent=\z@\relax
               \hbox to \Wdd@{\hfil}
               \hRule@\hRule@\hRule@\hRule@
               \vfil\hhrule\vfil}%
             \vrule\hskip-.4pt}
    }%
  \gridcount=0%
  \setbox\vGridbox@=\hbox{%
      \vbox{\offinterlineskip\parindent=0pt\hsize=0pt
         \vskip-.4pt\hrule%
         \hbox to \Wdd@{%
                 \vtop to \Htt@{\vfil}%
                 \vRule@\vRule@\vRule@\vRule@
                 \hfil\vvrule\hfil}%
         \hrule\vskip-.4pt}}%
  \wd\hGridbox@=0pt\ht\hGridbox@=0pt
  \wd\vGridbox@=0pt\ht\vGridbox@=0pt
  \hbox{\box\hGridbox@\box\vGridbox@}%
  }

 \def\LabelsGlobal{\def\LabGl@b{\global}}
 \def\LabelsLocal{\def\LabGl@b{}}
 \LabelsGlobal 

 \def\SetLabels#1\endSetLabels{%
   \LabGl@b\Labeltoks@={#1()\\}%
   }

 \LabGl@b\Labeltoks@={()\\}

 \def\ShowGrid{\LabGl@b\let\PlaceGrid@\PlaceGrid@@}
 \def\HideGrid{\LabGl@b\let\PlaceGrid@\relax}
 \def\Grids{\ShowGrid\LabGl@b\let\GridSwitch@\ShowGrid}
 \def\noGrids{\HideGrid\LabGl@b\let\GridSwitch@\HideGrid}

 \noGrids

 \def\bAdjust@@{%
     \setbox\auxLabelbox@=\hbox{\raise \dp\auxLabelbox@
            \box\auxLabelbox@}}
 \def\bAdjust@{\let\vAdjust@\bAdjust@@}

 \def\eAdjust@@{\dimen0=-.5\ht\auxLabelbox@
     \advance\dimen0 by .5\dp\auxLabelbox@
     \setbox\auxLabelbox@=
            \hbox{\raise\dimen0\box\auxLabelbox@}}
 \def\eAdjust@{\let\vAdjust@\eAdjust@@}

 \def\tAdjust@@{%
     \setbox\auxLabelbox@=\hbox{\raise-\ht\auxLabelbox@
            \box\auxLabelbox@}}
 \def\tAdjust@{\let\vAdjust@\tAdjust@@}

 \let\vAdjust@\relax

 \def\lAdjust@{\let\hAdjust@\rlap}
 \def\rAdjust@{\let\hAdjust@\llap}

 \let\hAdjust@\relax\let\vAdjust@\relax

 \def\FetchLabel@#1(#2)#3\\{%
     \IN@0#2@@\ifIN@
        \setbox0=\hbox{\ignorespaces#1#3\unskip}%
        \ifdim\wd0>0pt
           \ms@g{}%
           \ms@g{ !!! Bad label(s)? !!!}%
           \message{ #1(#2)#3}%
        \fi
        \def\LabelMole@##1\endFetchLabel@{%
            \IN@0()\\@##1@%
            \ifIN@\def\Temp@{\FetchLabel@##1\endFetchLabel@}%
            \else\def\Temp@{}%
            \fi
            \Temp@
           }%
     \else
       \ignorespaces#1\unskip
       \setbox\auxLabelbox@=%
         \hbox to 0pt{\hss\ignorespaces\hAdjust@
          {\ignorespaces#3\unskip}\hss}%
       \vAdjust@
       \let\hAdjust@\relax\let\vAdjust@\relax
       \AugmentLabelBox@@{#2}%
       \ht\Labelbox@=0pt\dp\Labelbox@=0pt
       \let\LabelMole@\FetchLabel@%
     \fi\LabelMole@}

 \newtoks\XYSep@ 
 \def\SetXYSeparator#1{%
     \IN@0#1@@\ifIN@\XYSep@{*}%
     \else
     \XYSep@{#1}%
     \fi
     }

 \SetXYSeparator*

 \def\AugmentLabelBox@@#1{%
     \IN@0\the\XYSep@ @#1@\ifIN@
       \SPLIT@0\the\XYSep@ @#1@%
       \setbox\Labelbox@=\hbox to 0pt{%
         \unhbox\Labelbox@
         \Shifted@@{\the\Initialtoks@\Wddd@}%
         {\the\Terminaltoks@\Httt@}%
         {\box\auxLabelbox@}}%
     \else
         \ms@g{}%
         \ms@g{ !!! Bad insertion point. !!!}%
         \message{ (#1\ this point was rejected.)}%
     \fi
    }

 \def\FetchOption@#1[#2]#3\endFetchOption@{%
    \def\temp{#1}
    \ifx\temp\empty
       \Edgect@=#2\relax
       \let\EdgGl@b\relax
       \SetEdges@
       \Cleaner@#3%
    \fi}

 \def\Cleaner@#1[@]{\Labeltoks@{#1}}
     
 \def\PlaceLabels@@{\mathsurround=0pt
     \def\Cr@{\\}%
     \let\L\lAdjust@\let\R\rAdjust@
     \let\B\bAdjust@\let\E\eAdjust@\let\T\tAdjust@
     \expandafter\FetchOption@\the\Labeltoks@[@]\endFetchOption@
     \Wddd@=\Wdd@ \Edgescale@\Wddd@ 
     \Httt@=\Htt@ \Edgescale@\Httt@
     \expandafter\FetchLabel@\the\Labeltoks@\endFetchLabel@
     \box\Labelbox@
     }%

 \let \PlaceLabels@\PlaceLabels@@

 \def\AffixLabels#1{\setbox\Coordinatebox@=\hbox{#1}%
      \Wdd@=\wd\Coordinatebox@ \Htt@=\ht\Coordinatebox@
      \advance\Htt@ \dp\Coordinatebox@
      \hbox{\copy\Coordinatebox@\kern-\Wdd@ 
           \Shifted@@{0pt}{-\dp\Coordinatebox@}%
           {\PlaceLabels@\PlaceGrid@}%
           \kern\Wdd@}%
      \GridSwitch@ 
      \LabGl@b\Labeltoks@{()\\}%
      }
 
   \let\wlog\LabelFigwlog@ld   
   \catcode`\@=\LabelFigCatAt  

                                By

              Raymond S\'eroul <A18645@FRCCSC21.BITNET>
                                and 
              Laurent Siebenmann <lcs@topo.math.u-psud.fr>
    
              VERSIONS: July 1991, Oct 1991, Jan 1992, July 1992

INTRODUCTION

      This labelling package is intended for TeX users who
rely on non-TeX sources for for their graphics inserts.  It
provides means for adding TeX labels to such inserts with a
minimum of fuss. 

       For most labels, TeX users have in the past found it
reasonably convenient to rely on non-TeX sources. Typical
occasions when an inescapable need for TeX labels seemed to
arise are

 (a) when the graphics program lacks certain exotic or complex
mathematical symbols

 (b) when the very highest typographical quality is wanted for the
labels

 (c) when labels included with the graphics fail to print, 
 and you cannot figure out why (cf. boxedeps.doc).  The labels
 provided by labelfig.tex are 100

       Since this package first appeared, many users, who in the
past scarcely dreamed of using TeX labels, have come to use
nothing but.  So it is now appropriate to add

Intoxication Warning:  TeX labels may be addictive and expensive. 

     If you have a fast preview you may disagree, and even find
that this package provides an agreeable paste-up environment; see
extra applications at end.

     Note to publishers: It is possible and convenient to ultimately
export the TeX labels produced by labelfig.tex to become an integral
part of the EPS file. This is often desired by a publisher who typically
uses an "upmarket" graphics or page layout program, with which the
staff is skilled in perfecting figures.  See Appendix I for
a recipe.

     The authors are grateful to Patrick Ion of Math Reviews for
helpful comments and encouragement.

BASIC INSTRUCTIONS

    After reading in the macro file using

preview or proof your figure with a coordinate grid printed on
top, by typing the following:

    \ShowGrid  
    \AffixLabels{<the graphics insertion>}

Here <the graphics insertion> is what you would type to insert
the graphics object alone without the grid.  This must provide
for the space around it. For example <the graphics insertion>
might well be \BoxedEPSF{MyFigure scaled 700} using the
boxedeps.tex macro package (from same source); this provides a
TeX box containing the encapsulated PostScript insert specified by
the file MyFigure. \AffixLabels{...} provides the grid (supposing
\ShowGrid is present) and later, once you have specified labels
using the grid, it will "tack on" the labels.

     The grid is a sort of (usually elongated) checkerboard of
ten rows and ten columns and its (internal) partitions are by
default numbered  .1, ... ,.9  both horizontally (X-coordinate
running left to right) and vertically (Y-coordinate running bottom
to top).  Thus the points enclosed by the grid correspond to the
points of the unit square in the cartesian "X-Y" plane, the lower
left corner corresponding to the origin (0,0).  By extrapolation,
the full page corresponds to a larger rectangle in the plane.

     These coordinates serve to position labels as follows.
Before the \AffixLabels{...} command type label specifications:

  \SetLabels
   (<X-coordinate>*<Y-coordinate>) <first label> \\
   .
   .
   .
   (<X-coordinate>*<Y-coordinate>)  <last label> \\
  \endSetLabels

Each row specifies one label and is terminated by \\.  In each
row, the position indicator comes first; it is written as a
standard cartesian point except that the X- and Y- coordinates
are separated by * rather than a comma because TeX allows a
comma as decimal point. There are no dimension units to specify
as the unit is the grid itself.

     By default, this cartesian point specifies where the middle
of the baseline of the label will be located.  However if you precede
the point by \L [or \R] the left [or right] edge of the baseline will
be located there. Similarly you may also precede the point by \T, \E,
or \B to vertically align the top equator or bottom of the label box
at the specified point.  This gives nine standard positions of
the label with respect to the insertion point --- corresponding to
the eight principle points of the compas and the center

                     \L\T     \T      \R\T

                     \L\E     \E      \R\E

                     \L\B     \B      \R\B

But this neglects the default "baseline" level of TeX,
giving potentially three more positions

                     \L    <no tag>   \R

For text, the baseline level is often the preferred. Its relation to
the others is variable. It will often coincide with the bottom level,
as happens for "X".  But it is often distinct, as for "g", in which
case you have in all 12 distinct positions rather than 9.

     It is convenient to think of this specification of label
position as attaching the label by a thumb-tack to the coordinate
grid. There are up to twelve positions of the thumb-tack on the
label, while the position of the thumb-tack on the coordinate grid is
arbitrary.  Normally, one choses the position of the thumb-tack on
the label to be the one that is the closest to the item being
labeled.  There are good reasons for this "rule of thumb":

   (a)  It facilitates correct positioning at first try.

   (b)  If the scale of the figure must be altered after labels
have been affixed, the labels have a good chance of remaining well
positioned.

   (c)  The visible grid need not extend beyond the "bounding box"
for the figure, because the best preferred position is always
(at least almost) within the bounding box .

The second reason is particularly important. Indeed it often
happens that scale has to be altered after labelling begins, in
order to either provide space for the labels, or to adjust
proportions between the labels and the figure.  (The size of labels
is unaffected by scaling.)

     Here is an artificial but self-contained test which uses
TeX rules to make a graphics object.

TEST

    Do not skip this!

 \def\FrameIt#1{\hbox{\vrule$\vcenter {\hrule\kern3pt%
             \hbox {\kern3pt #1\kern3pt}%
               \kern3pt\hrule}$\relax\vrule}}

 \def\Caption#1#2{\FrameIt{%
       \vtop {\hsize=#1\relax \parindent=0pt
         \leftskip=0pt \rightskip=0pt plus15pt
         \parfillskip=0pt
         \lineskip=1pt\baselineskip=0pt
         #2}}}

 \def\FirstQuadrant{\hbox to 100pt{\vrule\vbox to 100pt{%
        \hbox to 100pt{\hfil}\vfil\hrule}\hss}}

  \SetLabels
    \R(.5*.2) $\zeta\,\cdot$\\
    (.9*-.10) $\xi$\\
    \R(-.03*.9) $\eta$\\
    \T(.5*.9) \Caption{70pt}{%
          \it The norm of
          $g(\xi+i\eta)$ is indicated on
          contours of this invisible surface.}\\
  \endSetLabels

  \AffixLabels{\FirstQuadrant}

  \end

  Note that the coordinates to use for labels are indicated on the
edges of the grid (when visible) corresponding to the conventional
x- and y- axes of the Cartesian plane. By default the grid is
1-by-1. However, by the command \Edges{100}, you can change this
to 100-by-100 and many users find this alternative most
convenient. Place the command \Edges{...} in your style file (or
header) since its effect is is global. Other possible edge values
are 10 and 1000.

  If you use the command \Edges{...} at all, do so with care.  For
if you accidentally delete an \Edges{...} command your labels will
abruptly be badly misplaced and may logically but mysteriously
generate "dimension too big" errors under TeX and "off page" errors
under your driver.  

  You can dictate the edgescale for an individual figure by giving
the scale in brackets immediately after \SetLabels.  Thus, to
import into an article using say \Edge{100} a figure labelled using
another edgescale, say the original 1-by-1 default, you can use
\SetLabels[1]...\endSetLabels.

GETTING IT DOWN PAT

     Complicated labeling deserves the same respect as
complicated mathematics.  Do not expect it to come out perfect the
first time!  What is needed in either case is a mechanism to
repeatedly typeset troublesome pieces.

     One mechanism is always available.  One does complicated
labelling in a separate "test" file involving just the figure being
labelled;  a texpert will know how to \dump TeX's current state as
a temporary format that restarts rapidly at each retry.  Usually,
one then pastes the completed labelled figure back into the main
TeX file, but, of course, one can also \input it as an auxiliary
file.

     If you do not have a TeXpert at handy, here is a first
approximation to an efficient setup. By deletions reduce a copy
of your article to just a few lines before and after the figure.
Now label the figure, and finally, copy and paste the labelled
figure to the original article. Then copy the next figure to label
into this testbed and repeat. The TeXpert can improve the  speed
at which TeX starts up, by compiling a format specifically for
your article; just one caution: best NOT include in the format
ephemeral details of setup like \Set<mydriver>ArtSpecials (from
boxedeps.tex because this reads  figure dimensions which you may
change during your work session.

     An improved mechanism to repeatedly typeset troublesome
pieces is now available on the Macintosh; it is called LinoTeX;
see the same ftp sources.  It could be set up on many types
of computer.

     Before using labelfig.tex to attach labels to a graphics
object inserted using boxedeps.tex or BoxedArt.tex, make it a
firm rule to carefully adjust the bounding box using the trimming
commands of these packages, and also at least tentatively scale
and position the object. Beware of changing the grid inadvertently
after the labels have been positioned.  For example, correcting
the bounding box of a PostScript graphics object can foul up the
labels by changing the coordinate grid to which the labels are
attached. This is particularly true for the trimming  commands of
boxedeps.tex and BoxedArt.tex. However, as noted already, change
of scale is much less disruptive, and modest adjustments should be
well tolerated.

     Sometimes the labels protrude so far from the bounding box
of a figure that the figure has to be repositioned.  Best do this
by ad hoc spacing, say using \hglue and \vglue; altering the
bounding box would create a vicious circle.

     Remember that you are responsible for preventing labels
from overlapping. You are responsible for all label typography
including size and style. A label is really just about anything
that can be put in a TeX box. Note that spaces at the beginning
and end of labels will normally be suppressed; if you really want
them you must protect them with TeX braces.

     This package temporarily sets the \mathsurround parameter
of TeX to zero  while the labels are being affixed. This is done
because nonzero \mathsurround space would influence the position
of left and right aligned labels; then, when a texpert or printer
modifies mathsurround, diagram labeling might be disastrously
altered. There is a small price to pay involving labels that are
formatted as caption boxes including mathematics: you  may want or
need to specify an explicit mathsurround space within the caption
box; it will not influence anything outside.

     Those hostile to the use of * as separator between
the X and Y coordinates of label insertion points, are free to
impose another using \SetXYSeparator{<the new separator>}.  
Americans may prefer "," to "*" since they never use a 
comma as a decimal point; on the other hand, * may be more visible.

APPENDIX (I)  MERGING labelfig.tex LABELS INTO AN EPSF GRAPHICS OBJECT.

     As promised in the introduction, here is a recipe useful for
publishers. It works at least on Macintosh and at least for vectorized
graphics and Adobe type1 fonts.  (There is surely a similar recipe for
PCs under MSWindows.)

 (a)  Use boxedeps.tex utility to integrate the figure given by the eps
file, "x.eps" say, with a visible frame around it.  See
\ShowDisplacementBoxes command in boxedeps.tex.  To get precise results
automatically it is important to use the \Trim... commands of
boxedeps.tex making the "DisplacementBox" neatly fit the figure.

 (b)  Use the TeX printer driver and LaserWriter (versions >= 8.1.1) to
export to an EPSF the DVI page containing the integrated, labelled
figure. You now have an EPS file  "xx.eps"  that contains too much, and at
the wrong scale, and at wrong position.

 (c)  Convert the EPSF to an Adode Illustrator format EPSF using
the shareware utility called epsConvert by Sam Weiss
1993-- (currently $25).

 (d)  In Illustrator (or a compatible program), group the labels and the
"DisplacementBox"; copy them to the clipboard and paste them into "x.ps".
This step requires that all the label fonts be "visible to the Macintosh.

 (e)  Translate and scale the pasted group consisting of the labels plus
the "DisplacementBox" so as to make the "DisplacementBox" the bounding
box of (labelless) figure represented by "x.eps".  At this point the
labels will be correctly placed on the figure "x.eps".

 (f)  Ungroup and delete the "DisplacementBox".  The result is the
desired single EPS file, "x+.eps" say, It contains the original figure
plus its labels.  

     Using grouping and ungrouping appropriately in "x+.eps", a
publisher's staff can very efficiently improve label positions etc.

APPENDIX II)  SOME EXOTIC APPLICATIONS

     The grid of labelfig.tex is analogous to a light-table in
classical page makeup with wax or latex glue.  In principle, you
can use it to compose any page from its indivisible parts.  This
even has some of the artisanal charm of classical paste-up
provided you have a fast screen preview to make the process
"interactive".

     In practice labelfig.tex is a tool for nonstandard jobs.
Here are a few going beyond the labelling already discussed.

(I)  GRAPHICS INTEGRATION.

     This is accomplished by treating the imported graphics
objects as labels.  The underlying graphics object is then
typically an empty  \vbox to <dimension>{\vfill} in a TeX
\midinsert...\endinsert construction.  A label line
might be of the form

   (.1*.1) \\

The exact form of the special command varies from driver to
driver.  However, in the case of encapsulated PostScript graphics
(EPSF norm), by relying on boxedeps.tex, one can have the
following standard syntax (independant of driver  (see
boxedeps.doc for details.
  
  (.1*.1) \BoxedEPSF{MyFigure scaled <scale in mils>}\\

This may be slow since it requires TeX to read the PostScript
file to read bounding box using many complex macros.  So you
may want to try

  (.1*.1) \EPSFSpecial{MyFigure}{<scale in mils>}\\

which is fast and driver independant, but it squashes the
bounding box, normally to its lower left corner.

     Similarly for graphics of the Macintosh PICT norm ---
using BoxedArt.tex (same sources) in place of boxedeps.tex.

     This approach to integration is to be recommended when
one is assembling a composite graphics object.

 (II)  COMMUTATIVE DIAGRAM ENHANCEMENT

     Commutative diagrams or arrays of mathematical objects
connected by arrows of various sorts are common in mathematics.
The mathematical objects require the use of TeX.  Recently TeX
acquired a good collection of arrows of all slopes --- that of
LamSTeX --- plus pwerful macros to build the diagrams.

     However, even the LamSTeX collection is often
inadequate; it lacks for example double shafted arrows, dotted
arrows and curved arrows. Fortunately it is possible to produce
such arrows on an individual basis using sophisticated graphics
programs such as Illustrator and AldusFreehand (both serving
the EPSF norm) or using Metafont (with its public domain norm).
Since the creation of each new arrow is a work of love, you
probably want to limit the number of arrows by using LamSTeX
for most arrows. The 40K commutative diagram module of LamSTeX
has been adapted to work with AmSTeX and a copy may be posted
with LabelFig and related files. Unfortunately no one has yet
offered a version that works with Plain TeX or LaTeX.

       Suffice it here to say that when the exotic arrow has
been somehow imported into TeX, labelfig.tex treats it as a
label that one affixes to the commutative diagram.  Two other
steps will be treated in separate notes, namely the matter of
extracting the dimension specifications for the arrow and the
construction of the arrow --- for these steps are far from
unique and often depend intimately on your computer environment. 
Notes for the Macintosh-Textures-Illustrator combination are
found in the file ExoticArrows.doc.

 (III) NESTING 

Ingenuity pays off in exploiting labelfig.tex. One can
mix graphics and typography quite freely.  labelfig.tex is good
for freeform or overlapping arrangements, while boxedeps.tex (or
BoxedArt.tex) is best for regimented non-overlapping
arrangements --- and the two can be combined.

     The default behavior of labelfig.tex is not ideal 
for nesting objects, because to prevent trouble for beginners
the register for labels is globally cleared when \AffixLabels
concludes.  But there are switches available

      \LabelsGlobal      \LabelsLocal

which change this.  To understand this, extend the above test 
by something like:

 \LabelsLocal

 \SetLabels
    (.5*.5) AAA\\
 \endSetLabels

 {
 \SetLabels
    (.5*.5) ZZZ\\
 \endSetLabels
   \AffixLabels{\FirstQuadrant}
 }

   \AffixLabels{\FirstQuadrant}

     There are however potential pitfalls.  Neither
labelfig.tex nor boxedeps.tex has been tested under extreme
conditions. Problems may occur if their procedures are
indiscriminately nested. For boxedeps.tex (not labelfig.tex)
there is a precise cause for worry, namely many of its
variables are "global", which means that TeX braces will not
provide the protection one might expect.

COMMAND SUMMARY FOR labelfig.tex

  Here [...] means optional (one or zero)
       [...]* means any number of such constructs

  \SetLabels
    [[<P>](<X><Sep><Y>) <label> \\]*
  \endSetLabels
  \ShowGrid  
  \AffixLabels{<the figure>}

   --- <P> is tack position, one of eleven or empty
              order irrelevant

                   \L\T      \T      \R\T

                   \L\E      \E      \R\E

                     \L               \R

                   \L\B      \B      \R\B

   --- (<X><Sep><Y>) insertion point;
  <Sep> is separator, = * by default;
  \SetXYSeparator{<Sep>} changes it.
   <X> and <Y> are real numbers

  --- <label> a label to attach 

  --- <the figure> the figure to label 

  \GlobalLabels (default)     
  \LocalLabels  setting for nested constructs.

 \Grids makes ALL grids appear; \HideGrid then makes just next disappear.
 \noGrids returns to default.  The commands are always global.

 \GridLineWidth{<dimension>} adjusts width of grid lines. Default is very
small, to give "hairline" effect. If your grid lines are missing try
setting \GridLineWidth{1pt}.

 \Edges#1 globally changes the edge size of all grids to the numerical 
value #1, which must be 1, 10, 100, or 1000.  The default is 1.

VERSION HISTORY.
 --- Jan 1993: \Edges#1 and [??] option after \SetLabels
 --- July 1992: \Grids, \noGrids, \HideGrid;
       Gridlines become hairlines; \GridLineWidth{<dimension>}.
 --- Oct 1991, Jan 1992: \SetXYSeparator{<Sep>},  \LabelsGlobal,
       \LabelsLocal.
 --- July 1991: first release

Address for bugs and other feedback:

        Raymond S\'eroul
        IREM and Lab. de Typographie Informatise
        Univ. Rene Descartes
        Strasbourg

    Tel 33-88-41-63-45
    Email:  A18645@FRCCSC21.BITNET

        Laurent Siebenmann
        Mathematique, Bat. 425,
        Univ de Paris-Sud,
        91405-Orsay,
        France

    Tel 33-1-6941-7949; 
    Email: lcs@topo.math.u-psud.fr

\def\psfancypar#1#2{\begingroup\def\par{\endgraf\endgroup\lineskiplimit=0pt}
               \setbox2=\hbox{\large\sc #2}
               \newdimen\tmpht \tmpht \ht2 \advance\tmpht by \baselineskip
               \font\hhuge=Times-Bold at \tmpht
               \setbox1=\hbox{{\hhuge #1}}
               \count7=\tmpht \count8=\ht1
               \divide\count8 by 1000 \divide\count7 by \count8 
               \tmpht=.001\tmpht\multiply\tmpht by \count7 
               \font\hhuge=Times-Bold at \tmpht
               \setbox1=\hbox{{\hhuge #1}}
               \noindent
                \hangindent1.05\wd1
               \hangafter=-2 {\hskip-\hangindent
               \lower1\ht1\hbox{\raise1.0\ht2\copy1}%
                \kern-0\wd1}\copy2\lineskiplimit=-1000pt}

\newcommand{\E}{\mbox{{\rm E}}}

\def\boxit#1{\vbox{\hrule\hbox{\vrule\kern3pt
        \vbox{\kern3pt#1\kern3pt}\kern3pt\vrule}\hrule}}

\def\reals{ { {\rm  I \kern-0.15em R }  } }
\def\complex{ {\,{{\rm C} \kern-0.50em \raise0.20ex {  |}}\, }}

\def\Sigmabf{\hbox{$\bf \Sigma$}}

\def\sbf{{\bf s}}

\def\wbf{{\bf w}}

\def\ybf{{\bf y}}

\def\ybf{{\bf y}}

\def\Ibf{{\bf I}}

\def\Qbf{{\bf Q}}
\def\Rbf{{\bf R}}
\def\Sbf{{\bf S}}

\def\Ubf{{\bf U}}

\def\Wbf{{\bf W}}

\def\Ybf{{\bf Y}}

\def\Bc{{\cal B}}

\def\Fc{{\cal F}}

\def\Nc{{\cal N}}

\def\Xc{{\cal X}}

\def\be{\vskip .3cm \begin{equation}}
\def\ee{\end{equation} \vskip .4cm \noindent}
\def\defeq{{\stackrel{\Delta}{=}}}
\newcommand{\R}{\mbox{$\hat {\bf R}_{N}$}}

\def\Rxx{\Rbf_{\ssstyle X\kern-.1em X}}

\let\ssstyle=\scriptscriptstyle

\def\Kout{\setbox1=\hbox{\Huge\bf K}\hbox to
1.05\wd1{\hspace{.05\wd1}
}}
\def\Sout{\setbox1=\hbox{\Huge\bf S}\hbox to 1.05\wd1{\hspace{.05\wd1}
}}

\def\scalefig#1{\epsfxsize #1\textwidth}

\def\nn{\nonumber}
\def\defeq{\stackrel{\Delta}{=}}
\def\Ebb{{\mathbb E}}
\def\Rbb{{\mathbb R}}

\newcommand{\SNR}{\mbox{SNR}}

\newcommand{\beq}{\begin{equation}}
\newcommand{\eeq}{\end{equation}}

\title{Optimal and Suboptimal Detection of Gaussian Signals in Noise: Asymptotic Relative Efficiency}

\author{Youngchul Sung\supit{a}, Lang Tong\supit{a} and H. Vincent Poor\supit{$\dagger$b}
\skiplinehalf
\supit{a}School of Electrical and Computer Engineering, Cornell University, Ithaca, NY 14850\\
\supit{b}Dept. of Electrical Engineering, Princeton University,
Princeton, NJ 08544}

\authorinfo{$\dagger$Further author information: Send correspondence to
H. Vincent Poor. E-mail: poor@princeton.edu}

\begin{document}
  \maketitle

\begin{abstract}
The performance of Bayesian detection of Gaussian signals using
noisy observations is investigated via the error exponent for the
average error probability. Under unknown signal correlation
structure or limited processing capability it is reasonable to use
the simple quadratic detector that is optimal in the case of an
independent and identically distributed (i.i.d.) signal. Using the
large deviations principle, the performance of this detector
(which is suboptimal for non-i.i.d. signals) is compared with that
of the optimal detector for correlated signals via the asymptotic
relative efficiency defined as the ratio between sample sizes of
two detectors required for the same performance in the
large-sample-size regime.  The effects of SNR on the ARE are
investigated. It is shown that the asymptotic efficiency of the
simple quadratic detector relative to the optimal detector
converges to one as the SNR increases without bound for any
bounded spectrum, and that the simple quadratic detector performs
as well as the optimal detector for a wide range of the
correlation values at high SNR.
\end{abstract}


\keywords{Quadratic detector, error exponent, large deviations
principle, asymptotic relative efficiency (ARE)}

\section{INTRODUCTION}
\label{sec:intro}

We consider in this paper the optimal and suboptimal detection of
stationary Gaussian signals using noisy observations $y_i$ under a
Bayesian formulation. The corresponding null and alternative
hypotheses are given by
\begin{equation}  \label{eq:hypothesisscalar}
\begin{array}{lcl}
H_0 &: & y_i = w_i,  ~~~~i=1,2,\cdots, n,\\
H_1 &: & y_i = w_i+ \theta s_i, ~i=1,2,\cdots, n,\\
\end{array}
\end{equation}
where  $\{w_i\}$ is independent and identically distributed
(i.i.d.) $\Nc(0,\sigma^2)$ noise with a known variance $\sigma^2$,
$\theta$ is a nonnegative constant,  and $\{s_i\}$ is a zero-mean
unit-variance stationary Gaussian signal with spectrum
$f_s(\omega)$, independent of the noise $\{w_i\}$. The prior
probabilities  for the hypotheses are denoted by
\begin{equation}
\pi_0 \defeq \mbox{Pr}\{H_0\}, ~~\pi_1 \defeq \mbox{Pr} \{
H_1\}=1-\pi_0.
\end{equation}
Due to the stationarity of the signal, the signal-to-noise ratio
(SNR) for the observations is constant and is given by
\begin{equation}
\SNR = \frac{\theta^2}{\sigma^2}. \label{eq:SNRDEFscalar}
\end{equation}

Such a model arises, for example, in sensor networks (see, e.g.,
Sung {\em et
al.}\cite{Sung&Tong&Poor:05ICASSP,Sung&Tong&Poor:05IPSN}). For a
large sensor network deployed for the detection of stochastic
signals such as gases or particles in a fixed area, it is
reasonable to assume that the signal is random and that spatial
signal samples are correlated, while the measurement noise is
independent from sensor to sensor. Typically, the optimal detector
for (\ref{eq:hypothesisscalar}) is given in the form of a
quadratic detector that uses the correlation structure and
requires the joint processing of all signal samples. In general,
optimal detection using $n$ samples requires $O(n^2)$
multiplications and $O(n)$ memory size for storing past samples
except in some cases where recursive techniques are available.
\cite{Schweppe:65IT} These processing requirements may be
prohibitive in applications such as sensor network in which each
sensor node has stringent energy and storage constraints and the
number of nodes (or observation samples) is large.  Thus, one can
consider other detector structures with reduced complexity, e.g.,
simple quadratic detectors or banded-quadratic detectors.
\cite{Poor&Chang:85JASA,Chang&Poor:83CISS}

In this paper, we are interested in the asymptotic performance of
these detectors and the performance comparison between them using
the {\em asymptotic relative efficiency} (ARE) derived from the
large deviations principle (LDP)\cite{Chernoff:52AMS}.  Poor and
Chang investigated the performance of these detectors using
Pitman's ARE or asymptotic deflection
ratio.\cite{Poor&Chang:85JASA,Chang&Poor:83CISS,Capon:61IRE} While
ARE from the large deviations principle  is based on the law of
large numbers, Pitman's ARE relies on  convergence in distribution
(of the test statistics). Thus, these two ARE's do not necessarily
provide the same order for the performance of two detectors under
consideration, and Pitman's ARE generally provides more accurate
results than that of LDP in the low SNR
regime.\cite{vanderVaart:book} However, Pitman's ARE is based on
the asymptotic local scenario wherein the signal power decreases
to zero with a certain rate, i.e., typically $\theta$ in
(\ref{eq:hypothesisscalar}) decreases as $\frac{h}{\sqrt{n}}$ for
$h >0$ as the number $n$ of samples increases.  Thus, it does not
allow the performance comparison for a fixed signal-to-noise ratio
(SNR).   Poor and Chang considered the locally optimal detector as
the reference detector under the Neyman-Pearson formulation.  (The
efficacy\footnote{Pitman's ARE is expressed by the ratio of the
efficacy of one detector to that of the other.} of the optimal
quadratic detector is difficult to obtain since the amplitude
parameter $\theta$ is inseparable in the optimal test statistic,
as shown in (\ref{eq:optimalQbfn})).

The LDP for stationary Gaussian processes  is well-established.
\cite{Donsker&Varadhan:85CMP,Bryc&Dembo:97JTP,Bryc&Smolenski:93SPL,Bercu&Gamboa&Rouault:97SPA,Benitz&Bucklew:90IT}
Based on the result of Bryc and Dembo \cite{Bryc&Dembo:97JTP},
here we extend the work of Poor and
Chang\cite{Poor&Chang:85JASA,Chang&Poor:83CISS} and compare the
relative performance of several quadratic detectors using the ARE
from the LDP, focusing on the effects of SNR on the ARE with the
optimal detector as the reference detector under a Bayesian
formulation.

The paper is organized as follows.  In Section
\ref{sec:preliminaries}, some relevant results concerning the LDP
are presented. In Section \ref{sec:ARE}, the quadratic detectors
that we consider and the corresponding ARE are provided. In
Section \ref{sec:examples&numerical}, some numerical results are
presented for several examples of signal correlation, followed by
the conclusion in Section \ref{sec:conclusion}.

\section{Preliminaries}  \label{sec:preliminaries}

In this section, we present some definitions and results
concerning LDP relevant to the further development.

\vspace{0.5em}
\begin{definition}[Large deviations principle \cite{Dembo&Zeitouni:book}]
Let $\{P_n\}$ be a sequence of probability distributions defined
on $(\Xc, \Fc)$.  $\{P_n\}$ is said to satisfy the {large
deviation principle} with a rate function $I: \Xc \rightarrow
[0,\infty]$ if
\begin{itemize}
\item the level sets $I^{-1}([0,c])$ are compact for all $c <
\infty$,
 \item
\[\limsup_{n\rightarrow\infty}\frac{1}{n}\log{P_n(C)}\le
- \inf_{x\in C}{I(x)}~~~\forall ~{\mathrm closed}~C\in \Fc,\]
 \item and \[\liminf_{n\rightarrow\infty}\frac{1}{n}\log{P_n(O)}\ge
 -\inf_{x\in O}{I(x)}~~~\forall  ~{\mathrm open}~O\in \Fc.\]
\end{itemize}
\end{definition}
\vspace{1em}

For the probability distributions governing a sequence of sample
means the LDP is given by Cr\'amer's theorem, and its extension to
general sequences of random variables is provided by the
G\"artner-Ellis theorem based on the convergence of cumulant
generating functions.\cite{Dembo&Zeitouni:book,Hollander:book} In
particular, for the sequence of quadratic functionals of Gaussian
processes the rate function is derived by Bryc and Dembo
\cite{Bryc&Dembo:97JTP} circumventing difficulties  in applying
the G\"artner-Ellis theorem to this problem, which is summarized
in the following theorem.

\vspace{0.5em}
\begin{theorem}[Bryc and Dembo \cite{Bryc&Dembo:97JTP}]
\label{theo:Bryc&Dembo} Let $\{Y_i, -\infty < i < \infty \}$ be a
(real-valued) zero-mean stationary Gaussian process with bounded
spectral density function $S_y(\omega)$ defined as
\begin{equation}\label{eq:spectraldensity}
 S_y(\omega)= \sum_{k=-\infty}^\infty
\Ebb \{ Y_0 Y_{k}\} e^{-jk\omega}
\end{equation}
with essential supremum $M$. Let a random variable $Z_n \defeq
\{\frac{1}{n}\sum_{i=1}^n Y_i^2\}$ and $P_n$ be the distribution
of $Z_n$, i.e., $P_n(S)
\defeq \mbox{Pr}\{ Z_n \in S\}$ for $S \in \Bc(\mathbb{R})$. Then,
$\{P_n\}$ satisfies the LDP with a rate function
\begin{equation} \label{eq:ratefunction}
I(z) = \sup_{-\infty < t < \frac{1}{2M}} [zt - \Lambda(t) ],
\end{equation}
where \begin{equation} \label{eq:MGFGaussian} \Lambda(t) = -
\frac{1}{4\pi} \int_{0}^{2\pi} \log ( 1 - 2 t S_y(\omega)) d\omega
\end{equation}
\end{theorem}
\vspace{0.5em}

\vspace{0.5em}
\begin{lemma}[Bryc and Dembo \cite{Bryc&Dembo:97JTP}]
\label{lemma:Bryc&Dembo} Suppose $\Ybf=[Y_1,\cdots,Y_n]^ T$ is a
real-valued zero-mean Gaussian vector with the covariance matrix
$\Sigmabf$ and let $\Wbf$ be a symmetric real-valued $n\times n$
matrix. Then, with $\lambda_1,\cdots, \lambda_n$ the eigenvalues
of the matrix $\Wbf \Sigmabf$ we have
\[
\log \Ebb e^{s \Ybf^T\Wbf\Ybf}= -\frac{1}{2} \sum_{i=1}^n \log
(1-2s \lambda_i)
\]
for all $s\in \mathbb{C}$ s.t. $\max_i \{Re(s) \lambda_i\} < 1/2$.
Furthermore, $\log \Ebb e^{t\Ybf^T\Wbf \Ybf}=\infty$ for all $t
\in \Rbb$ s.t. $\max_i \{t \lambda_i\} \ge 1/2$.
\end{lemma}
\vspace{0.5em}

Another useful result concerns the asymptotic distribution of the
eigenvalues of a Toeplitz matrix, which is summarized in the
following theorem.

 \vspace{0.5em}
\begin{theorem}[Grenander and Szeg\"o \cite{Grenander&Szego:book}] \label{theo:ToeplitzDT}
Let $S_y(\omega)$ be the spectrum of $\{Y_i\}$, defined as
(\ref{eq:spectraldensity}), with finite lower and upper bounds
denoted by $m$ and $M$, respectively. Let $\Sigmabf_{y,n}$ be a
covariance matrix defined as
\begin{equation}
\Sigmabf_{y,n}=[\Ebb \{Y_i Y_j\}]_{i,j=1}^n
\end{equation}
and $\lambda_1^{(n)}, \cdots, \lambda_n^{(n)}$ be the eigenvalues
of $\Sigmabf_{y,n}$. Then, for any continuous function $h:[m,M]
\rightarrow {\mathbb R}$, we have
\begin{equation}
\lim_{n\rightarrow \infty} \frac{1}{n}\sum_{i=1}^n
h(\lambda_i^{(n)}) = \frac{1}{2\pi}\int_0^{2\pi}
h(S_y(\omega))d\omega.
\end{equation}
\end{theorem}
\vspace{0.5em}

\section{Asymptotic Relative Efficiency}  \label{sec:ARE}

In this section, we  present the classes of detectors that we
consider and their corresponding  rate functions.  By stacking the
observations and corresponding signals and noises, the hypotheses
(\ref{eq:hypothesisscalar}) can be rewritten in  vector form as
\begin{equation}  \label{eq:hypothesisvecnote}
\begin{array}{lcl}
H_0 &: & \ybf_n = \wbf_n,\\
H_1 &: & \ybf_n = \wbf_n+ \theta \sbf_n,\\
\end{array}
\end{equation}
where
\begin{equation} \nonumber
\ybf_n \defeq [y_0, \cdots, y_n]^T, ~~\sbf_n \defeq [s_0, \cdots,
s_n]^T, ~~\wbf_n \defeq [w_0, \cdots, w_n]^T,
\end{equation}
and the noise vector $\wbf_n \sim \Nc({\mathbf 0},\sigma^2 \Ibf)$,
 $\sbf_n \sim \Nc({\mathbf 0},\Sigmabf_{s,n})$, and $\ybf_n$
 has distribution $\Nc(\mathbf{0},\Sigmabf_{j,n})$ for hypothesis
 $j$ ($j=0,1$) where
\begin{equation} \label{eq:covariancemtx}
\Sigmabf_{0,n} = \sigma^2 \Ibf,~~~ \Sigmabf_{1.n} =\sigma^2 \Ibf +
\theta^2\Sigmabf_{s,n}.
\end{equation}
For convenience, we further assume equal prior probabilities,
i.e.,
\begin{equation}
\pi_0 = \pi_1=\frac{1}{2}.
\end{equation}
Then, the optimal detector for (\ref{eq:hypothesisvecnote}) is
given by the maximum {\em a posteriori} probability detector:
\begin{equation} \label{eq:optimaldetector}
\delta_o(\ybf_n) = \left\{
\begin{array}{cccc}
1,  & & \frac{1}{n} \log L_n(\ybf_n)  \ge \tau=0, \\
0,  & & \mbox{otherwise},  \\
\end{array}
\right.
\end{equation}
where
\begin{equation} \label{eq:LROPT}
L_n(\ybf_n) =  \left(
\frac{|\Sigmabf_{0,n}|}{|\Sigmabf_{1,n}|}\right)^{1/2}e^{\frac{1}{2}
\ybf_n^T \Qbf_n \ybf_n},
\end{equation}
and
\begin{equation}  \label{eq:optimalQbfn}
\Qbf_n =\Sigmabf_{0,n}^{-1} - \Sigmabf_{1,n}^{-1}= \sigma^{-2}\Ibf
- (\sigma^2\Ibf +\theta^2\Sigmabf_{s,n})^{-1}.
\end{equation}

Since the calculation of the likelihood ratio  requires the
product of all observations, the optimal detector typically
requires $O(n^2)$ multiplications and $O(n)$ memory for the
storage of the previous samples \cite{Poor&Chang:85JASA}.  Next,
we consider a simple quadratic detector obtained by neglecting the
signal correlation, i.e., $\Sigmabf_{s,n}\equiv \Ibf$, and  it is
given by
\begin{equation} \label{eq:simplequaddetector}
\delta_{sq}(\ybf_n) = \left\{
\begin{array}{cccc}
1,  & & \frac{1}{n} \log \left[ \left( \frac{\sigma^{2n}}{(\sigma^2+\theta^2)^n} \right)^{1/2} e^{\frac{1}{2} \ybf_n^T \hat{\Qbf}_{n} \ybf_n}\right]  \ge 0, \\
0,  & & \mbox{otherwise},  \\
\end{array}
\right.
\end{equation}
where
\begin{equation}
\hat{\Qbf}_n = \frac{\theta^2}{\sigma^2(\sigma^2+\theta^2)} \Ibf.
\end{equation}
The test statistic in this case can be rewritten as
\begin{equation}  \label{eq:teststatisticsq}
T_{sq,n}= \frac{1}{2}\log \frac{\sigma^2}{\sigma^2+\theta^2} +
\frac{\theta^2}{2n\sigma^2(\sigma^2+\theta^2)}\sum_{i=1}^n y_i^2.
\end{equation}
Thus, the simple quadratic detector requires $O(n)$
multiplications and one storage for accumulation.

We also consider a banded-quadratic detector structure which has
intermediate complexity between the optimal and the simple
quadratic detector, similar to that considered by Poor and
Chang.\cite{Poor&Chang:85JASA,Chang&Poor:83CISS} Since the
determinants of the two matrices $\Sigmabf_{0,n}$ and
$\Sigmabf_{1,n}$ can be computed off line for the optimal detector
(\ref{eq:optimaldetector}, \ref{eq:LROPT}) when the signal
correlation structure is known beforehand, the main complexity
results from the calculation of the quadratic term based on
observations. Thus, a class of detectors with intermediate
complexity is given by
\begin{equation} \label{eq:bandeddetector}
\delta_{b,m}(\ybf_n) = \left\{
\begin{array}{cccc}
1,  & & \frac{1}{n} \log
{L}_n^{(b,m)}(\ybf_n)  \ge 0, \\
0,  & & \mbox{otherwise},  \\
\end{array}
\right.
\end{equation}
where
\begin{equation} \label{eq:bandedquadLn}
{L}_n^{(b,m)}(\ybf_n) =  \left(
\frac{|\Sigmabf_{0,n}|}{|\Sigmabf_{1,n}|}\right)^{1/2}e^{\frac{1}{2}
\ybf_n^T \tilde{\Qbf}_n^{(m)} \ybf_n},
\end{equation}
and $\tilde{\Qbf}_n^{(m)}$ is a banded $n\times n$ symmetric
positive-definite Toeplitz matrix with bandwidth $(2m+1)$, i.e.,
\begin{equation} \label{eq:bandedquadtQbf}
\tilde{\Qbf}_n^{(m)} = \left[
\begin{array}{ccccccc}
  b_0    & b_1     & \cdots & b_m    & 0    & \cdots & \cdots  \\
  b_1    & b_0     & b_1    & \ddots & b_m &  0  & \cdots \\
  \vdots & b_1     & b_0    & b_1    &  \ddots & b_m   &0\\
  b_m    &  \ddots & \ddots & \ddots &  \ddots    & \ddots&b_m\\
  0      & b_m     &  \ddots&   b_1  &  b_0    & b_1   &\vdots\\
  \vdots & 0       & b_{m}  & \ddots &  b_1    & b_0  & b_1\\
  \vdots & \vdots  & 0      & b_m &  \cdots    & b_1  & b_0\\
  \end{array}
\right].
\end{equation}
Here, the values of $b_l, ~l=0,1,\cdots, m~ (b_{-l}=b_l)$ need to
be properly determined for optimal performance.  Let the
discrete-time Fourier transform of the finite sequence
$\{b_{-m},b_{-m+1},\cdots, b_{-1},b_0,b_1,\cdots, b_m\}$ be
$g_m(\omega)$, i.e.,
\begin{equation}  \label{eq:bandedDTFT}
g_m(\omega) = b_0 + 2 \sum_{l=1}^m b_l \cos(l\omega), ~~0 \le
\omega \le 2\pi.
\end{equation}
Then, the eigenvalues of $\tilde{\Qbf}_n^{(m)}$ converge to
 uniform samples, $\{g_m(\frac{2\pi k}{n})\}_{k=0,1,\cdots,n-1}$,
of $g_m(\omega)$ as $n$ increases since $\tilde{\Qbf}_n^{(m)}$ is
Toeplitz \cite{Grenander&Szego:book}.

\subsection{Error Exponent and ARE}

The false alarm probability, $\alpha_n^{(\delta)}$, and the miss
probability, $\beta_n^{(\delta)}$, for a particular detector
$\delta$ are defined as
\begin{eqnarray}
\alpha_n^{(\delta)} &\defeq& \mbox{Pr}\{\delta(\ybf_n)=1|H_0\},  \\
\beta_n^{(\delta)} &\defeq& \mbox{Pr}\{ \delta(\ybf_n)=0|H_1\}.
\end{eqnarray}
In general, these probabilities decay exponentially as $n$
increases without bound, and the decay rate is given by Theorem
\ref{theo:Bryc&Dembo}. Thus, we have
\begin{eqnarray}
E_0(\delta)&\defeq&-\lim_{n\rightarrow \infty} \frac{1}{n} \log \alpha_n^{(\delta)} =  \inf_{z \in [0,\infty)} I_0^{(\delta)}(z),\\
E_1(\delta)&\defeq&-\lim_{n\rightarrow \infty}\frac{1}{n} \log
\beta_n^{(\delta)} = \inf_{z \in (-\infty, 0)} I_1^{(\delta)}(z),
\end{eqnarray}
where $I_j{(\delta)}(z)$, $j=0,1$, is defined as
(\ref{eq:ratefunction}) with limiting cumulant moment generating
function $\Lambda_j^{(\delta)}(t)$ corresponding to the considered
detector and hypothesis.  The error exponent or the exponential
decay rate of the average error probability for the detector
$\delta$ is given by
\begin{eqnarray}
E(\delta) &\defeq& \lim_{n\rightarrow \infty} -\frac{1}{n} \log
P_{e,n}^{(\delta)}=\lim_{n\rightarrow \infty} -\frac{1}{n}
\log(\pi_0
\alpha_n^{(\delta)} + \pi_1 \beta_n^{(\delta)}),\nn\\
&=&  \min \{ E_0(\delta), E_1(\delta) \}.  \label{eq:minE0E1}
\end{eqnarray}
Hence, we have asymptotically
\begin{equation}  \label{eq:asympapprox}
P_{e,n}^{(\delta)}\sim e^{-nE(\delta)}.
\end{equation}
Eq. (\ref{eq:asympapprox}) provides an asymptotic criterion for
the comparison of two detectors \cite{Chernoff:52AMS}. The
 efficiency of $\{\delta_{1}\}$ relative to $\{\delta_{2}\}$ for
sample size $n$ is defined to the ratio $n_2/n$, where $n_2$ is
the smallest number of samples such that
$P_{e,n_2}^{(\delta_{2})}\le P_{e,n}^{(\delta_{1})}$
\cite{Poor:book}. Thus, the asymptotic  efficiency of a detector
$\delta_1$ relative to another detector $\delta_2$ from the LDP is
defined as the ratio between the two error exponents:
\begin{equation}  \label{eq:AREdef}
\mbox{ARE}_{\delta_1,\delta_2} \defeq
\frac{E(\delta_1)}{E(\delta_2)}.
\end{equation}

Now let us consider the rate function for each detector under
consideration.  For the simple quadratic detector
$\delta_{sq}(\ybf_n)$ the calculation of the rate function under
each hypothesis is straightforward from Theorem
\ref{theo:Bryc&Dembo}. Applying Theorem \ref{theo:Bryc&Dembo} to
(\ref{eq:teststatisticsq}), we have
\begin{eqnarray}
\Lambda_0^{(\delta_{sq})}(t) &=&
\frac{t}{2}\log\frac{\sigma^2}{\sigma^2+\theta^2}-\frac{1}{2}\log\left(1-t\frac{\theta^2}{\sigma^2+\theta^2}\right),\label{eq:Lam0sq}\\
\Lambda_1^{(\delta_{sq})}(t)
&=&\frac{t}{2}\log\frac{\sigma^2}{\sigma^2+\theta^2}-\frac{1}{4\pi}\int_0^{2\pi}\log\left(
1-t\frac{\theta^2(\sigma^2+\theta^2f_s(\omega))}{\sigma^2(\sigma^2+\theta^2)}\right)d\omega,\label{eq:Lam1SQt}
\end{eqnarray}
where $f_s(\omega)$ is the spectrum of the signal and the range of
$t$ is defined for each case so that the term in the logarithmic
function is strictly positive.

For the optimal detector $\delta_o(\cdot)$ the test statistic is
given by
\begin{equation}
T_{o,n} = \frac{1}{n} \log \left[\left(
\frac{|\Sigmabf_{0,n}|}{|\Sigmabf_{1,n}|}\right)^{1/2}e^{\frac{1}{2}
\ybf_n^T \Qbf_n \ybf_n}\right].
\end{equation}
In this case the rate function is obtained by a whitening
transform. Let the eigendecomposition of the signal covariance
matrix $\Sigmabf_{s,n}$ be
\begin{equation}
\Sigmabf_{s,n}=\Ubf \Lambda \Ubf^T = \Ubf
\mbox{diag}(\lambda_1^{(n)},\cdots, \lambda_n^{(n)})\Ubf^T,
\end{equation}
where $\Ubf$ is an orthogonal matrix. Then, the eigendecomposition
of $\Qbf_n$ is given by
\begin{eqnarray}
\Qbf_n &=& \Ubf \Sbf \Ubf^T, \\
 &=& \Ubf
\mbox{diag}\left(\frac{\theta^2\lambda_1^{(n)}}{\sigma^2(\sigma^2+\theta^2\lambda_1^{(n)})},\cdots,
\frac{\theta^2\lambda_n^{(n)}}{\sigma^2(\sigma^2+\theta^2\lambda_n^{(n)})}\right)\Ubf^T,\nn
\end{eqnarray}
and
\begin{equation}
\ybf_n^T \Qbf_n \ybf_n = ||\bar{\ybf}_n||^2 = \sum_{i=1}^n
\bar{y}_i^2,
\end{equation}
where $\bar{\ybf}_n = \Sbf^{1/2}\Ubf^T\ybf_n$ and $\bar{y}_i$ is
the $i^{\rm th}$ element of $\bar{\ybf}_n$. Thus, the test
statistic is given by
\begin{eqnarray*}
T_{o,n} &=& \frac{1}{2n}\sum_{i=1}^n \log
\frac{\sigma^2}{\sigma^2+\theta^2\lambda_i^{(n)}} +
    \frac{1}{2n}\sum_{i=1}^n \bar{y}_i^2.
\end{eqnarray*}
By Theorems \ref{theo:Bryc&Dembo} and \ref{theo:ToeplitzDT}, we
have
\begin{eqnarray}
\Lambda_0^{(\delta_{o})}(t) &=& \frac{t}{4\pi}\int_0^{2\pi}
\log\frac{\sigma^2}{\sigma^2+\theta^2f_s(\omega)}d\omega-\frac{1}{4\pi}\int_0^{2\pi}\log\left(
1-t\frac{\theta^2f_s(\omega)}{\sigma^2+\theta^2f_s(\omega)}\right)d\omega,\\
\Lambda_1^{(\delta_{o})}(t)&=&\frac{t}{4\pi}\int_0^{2\pi}
\log\frac{\sigma^2}{\sigma^2+\theta^2f_s(\omega)}d\omega-\frac{1}{4\pi}\int_0^{2\pi}\log\left(
1-t\frac{\theta^2f_s(\omega)}{\sigma^2}\right)d\omega
\label{eq:Lam1OPTt}.
\end{eqnarray}
The range of $t$ is again defined for each case so that the term
in the logarithmic function is positive. For the optimal case the
rate function for the quadratic part has also been derived by
several other authors, e.g., Chamberland\cite{Chamberland:thesis}.

For the banded quadratic detector the test statistic is given by
\begin{eqnarray}  \label{eq:bandedteststatsumform}
T_{b,n}^{(m)} &=& \frac{1}{n} \log \left[\left(
\frac{|\Sigmabf_{0,n}|}{|\Sigmabf_{1,n}|}\right)^{1/2}e^{\frac{1}{2}
\ybf_n^T \tilde{\Qbf}_n^{(m)} \ybf_n}\right]
=\frac{1}{2n}\sum_{i=1}^n \log
\frac{\sigma^2}{\sigma^2+\theta^2\lambda_i^{(n)}} +
    \frac{1}{2n} \ybf_n^T \tilde{\Qbf}_n^{(m)} \ybf_n,
\end{eqnarray}
where $\tilde{\Qbf}_n^{(m)}$ is defined  in
(\ref{eq:bandedquadtQbf}).  By Lemma \ref{lemma:Bryc&Dembo}, the
cumulant generating function for the quadratic part under the
hypothesis $j$ is given by
\begin{equation}
\log \Ebb_j \{e^{ t\frac{1}{2}\ybf_n^T \tilde{\Qbf}_n^{(m)}
\ybf_n}\} = -\frac{1}{2} \sum_{i=1}^n \log \left(1  -  t
\tilde{\lambda}_{ij}^{(n)}\right), ~~~j=0,1,
\end{equation}
for all $t < 1/(\max_i \tilde{\lambda}_{ij}^{(n)})$, where
$\{\tilde{\lambda}_{ij}^{(n)},i=1,\cdots,n\}$ are the eigenvalues
of $\tilde{\Qbf}_n^{(m)}\Sigmabf_{j,n}$,  and $\Sigmabf_{j,n}
~(j=0,1)$ is defined in (\ref{eq:covariancemtx}).  Because of the
Toeplitz structure of $\tilde{\Qbf}_n^{(m)}$ and $\Sigmabf_{j,n}$,
it follows that\cite{Grenander&Szego:book}
\begin{equation}  \label{eq:ToeplitzProduct}
\lim_{n\rightarrow \infty} \frac{1}{n} \sum_{i=1}^n h(
\tilde{\lambda}_{ij}^{(n)})= \frac{1}{2\pi} \int_0^{2\pi}
h(g_m(\omega) f_j (\omega)) d\omega
\end{equation}
for any continuous function $h(\cdot)$, where $f_j(\omega)$ is the
spectrum of the observation process $\{y_i\}$ under the hypothesis
$j~(j=0,1)$. Thus, the rate function for the banded-quadratic
detector is
 given by
\begin{eqnarray}
&& \Lambda_0^{(\delta_{b,m})}(t) = \frac{t}{4\pi}\int_0^{2\pi}
\log\frac{\sigma^2}{\sigma^2+\theta^2f_s(\omega)}d\omega-\frac{1}{4\pi}\int_0^{2\pi}\log\left(
1-t \sigma^2 g_m(\omega)\right)d\omega,\label{eq:bandedLam0}\\
&&\Lambda_1^{(\delta_{b,m})}(t)=\frac{t}{4\pi}\int_0^{2\pi}
\log\frac{\sigma^2}{\sigma^2+\theta^2f_s(\omega)}d\omega-\frac{1}{4\pi}\int_0^{2\pi}\log\left(
1-t (\sigma^2+\theta^2f_s(\omega))g_m(\omega)
\right)d\omega,\label{eq:bandedLam1}
\end{eqnarray}
 where $g_m(\omega)$ is defined in (\ref{eq:bandedDTFT}) and the
range of $t$ is defined properly in each case.

\section{Examples and Numerical Results}
\label{sec:examples&numerical}

We now consider some signal examples and investigate the relative
performance of the detectors in the previous section as a function
of various parameters such as correlation strength and SNR via the
asymptotic relative efficiency defined in (\ref{eq:AREdef}).
 In particular, we consider Gauss-Markov signals and triangularly correlated
signals. Except for some simple cases such as autoregressive
signals without additive noise \cite{Bryc&Smolenski:93SPL} it is
difficult to obtain closed-form expressions for the rate functions
in the previous section.  Thus, we evaluate the rate and ARE by
numerical evaluation of the error exponent.

\subsection{Gauss-Markov Signal}

We first consider the stationary Gauss-Markov signal for which the
correlation is given by
\begin{equation}
\Ebb\{S_0 S_k\} = a^{|k|}, ~k=0, \pm 1, \pm 2, \cdots, (0 \le a <
1),
\end{equation}
and the spectrum is given by the Poisson kernel:
 \begin{equation} \label{eq:spectrumGM}
  f_s(\omega) = \frac{1-a^2}{1-2a\cos \omega + a^2}.
 \end{equation}

\begin{figure}[htbp]
\centerline{ \SetLabels
\L(0.24*-0.08) (a) \\
\L(0.775*-0.08) (b) \\
\endSetLabels
\leavevmode 
\strut\AffixLabels{
\scalefig{0.407}\epsfbox{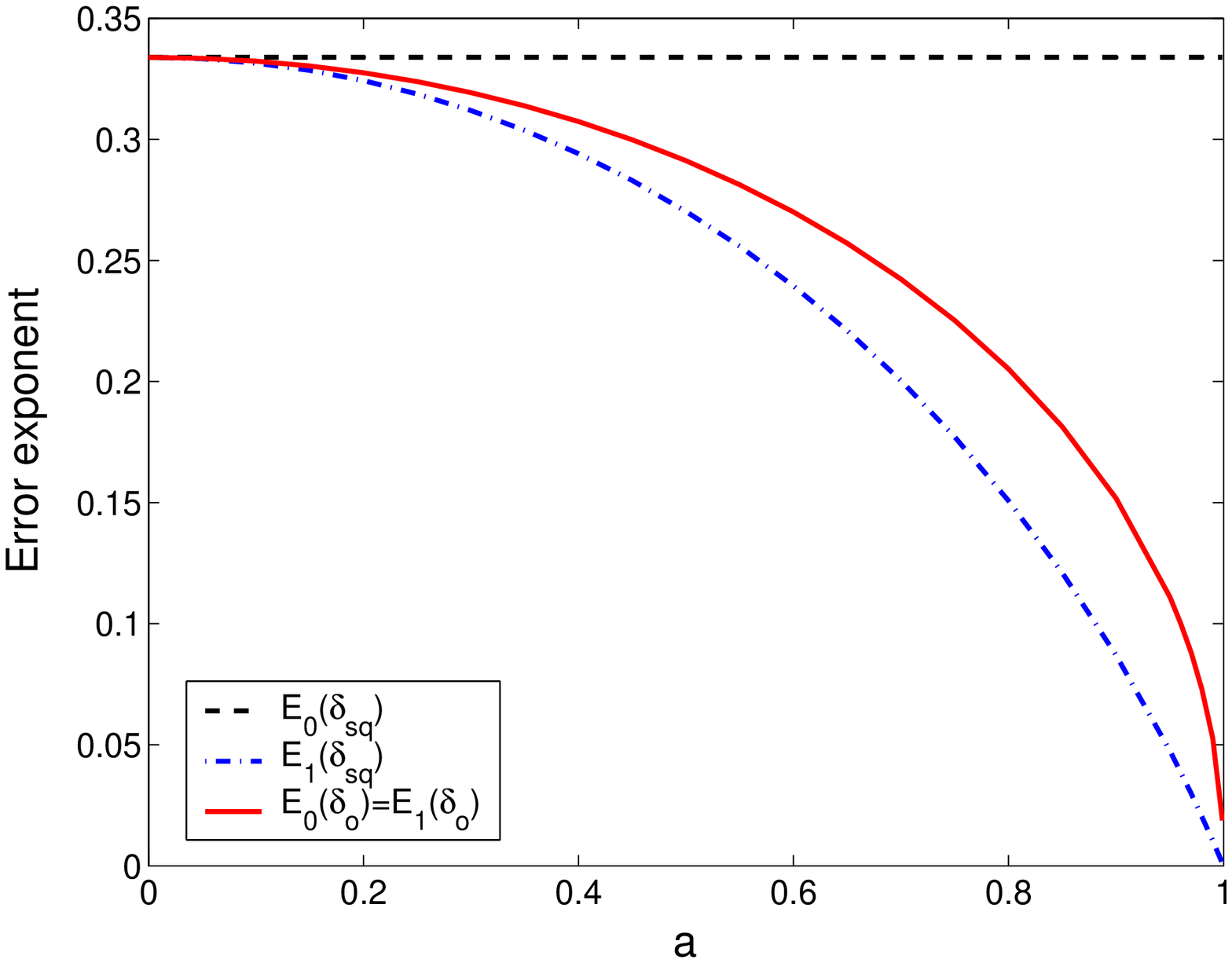}
\hspace{1cm} \scalefig{0.4}\epsfbox{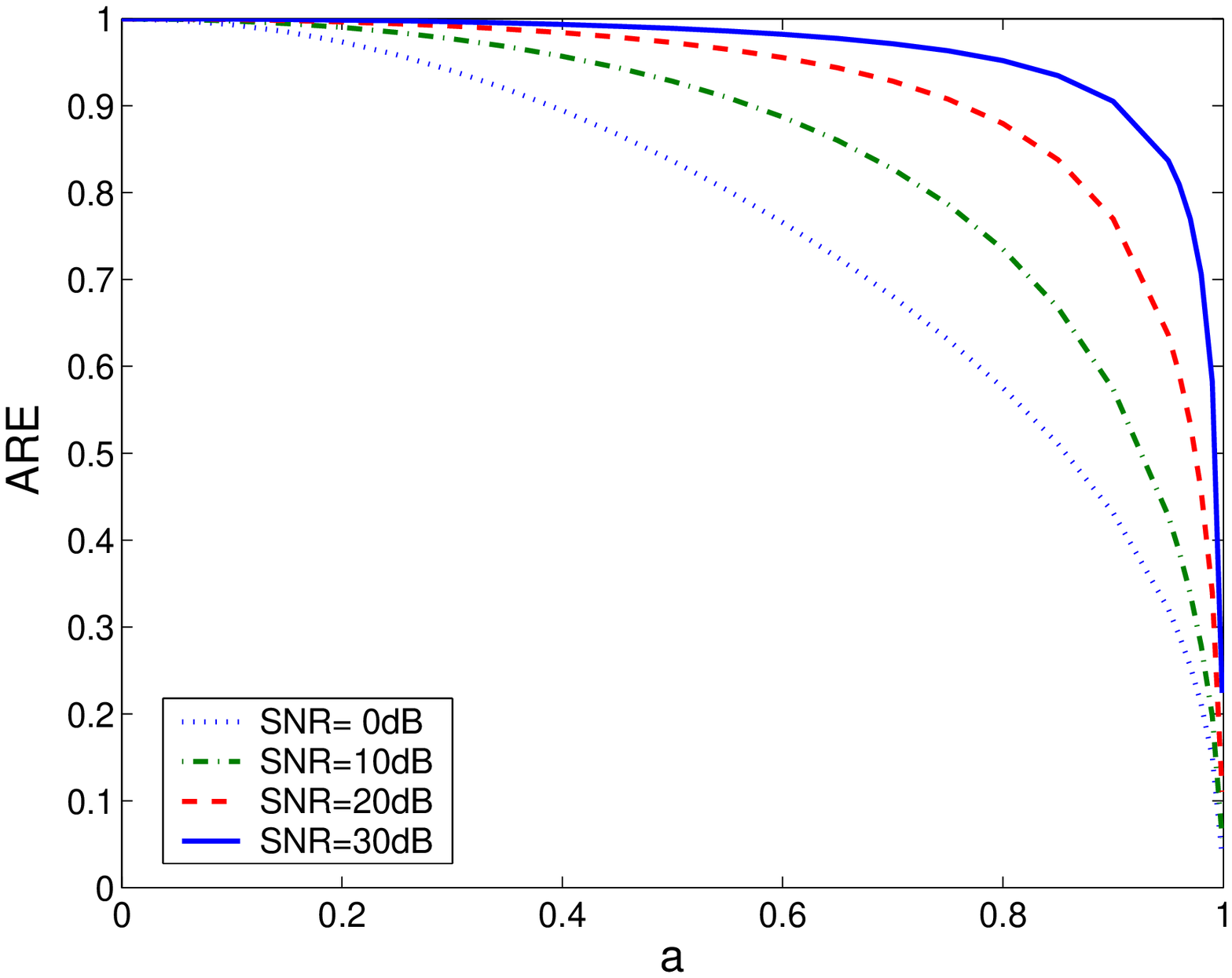}
} }
 \vspace{0.1cm} \caption{Optimal and simple quadratic detectors (Gauss-Markov signal): (a) error exponent,
$E_j(\delta)$, $j=0,1$, as a function of correlation strength $a$
(SNR=10dB) and (b) ARE as a function of correlation strength $a$
for SNR = 0, 10, 20, 30 dB.} \label{fig:RateComparisonGM}
\end{figure}

Fig. \ref{fig:RateComparisonGM} (a) shows the error exponent for
the false alarm and miss probabilities for the optimal and simple
quadratic detectors as a function of the correlation strength $a$
at 10 dB SNR.  It is seen that the error exponent
$E_0(\delta_{sq})$ for the false alarm probability of the simple
quadratic detector is independent of the correlation strength and
is equal to the maximum value of the error exponent of the optimal
detector achieved by independent signal\footnote{This is not the
case when the SNR is low. At low SNR the maximum value of the
error exponent for the optimal detection is achieved at some
correlation value $0 < a < 1$
\cite{Sung&Tong&Poor:05ISIT}.}($a=0$). This is easily seen by the
logarithmic generating function (\ref{eq:Lam0sq})  which does not
depend on the signal spectrum.
However, the error exponent $E_1(\delta_{sq})$ for the miss
probability is less than that of the false alarm probability for
all values of $a$, and decreases to zero as the signal correlation
becomes strong ($a \rightarrow 1$). Thus, the error exponent for
the average error probability is determined by that of the miss
probability for the simple quadratic detector. On the other hand,
the error exponents for the false alarm and miss probabilities are
the same, i.e., $E_0(\delta_o)= E_1(\delta_o)$, for the optimal
detector with equal prior probabilities, i.e., zero threshold in
(\ref{eq:optimaldetector}). In this case, the minimum in
(\ref{eq:minE0E1}) is attained and the error exponent is the
Chernoff information between the two distributions under the
hypotheses (\ref{eq:hypothesisvecnote}). Note that the error
exponent for the miss probability of the simple quadratic detector
is
 smaller than that of the optimal detector for $0 < a <
1$. So, the error exponent of the simple quadratic detector is
smaller than that of the optimal detector even if the simple
quadratic detector performs better than the optimal detector for
the false alarm probability. From the detector structure
(\ref{eq:teststatisticsq}) one can see that the simple quadratic
detector is optimized for the detection of the false alarm event
regardless of the signal correlation, thereby sacrificing the
performance for correct detection, while the optimal detector
optimizes the test statistic so that it can perform equally well
for both of the false alarm and miss events.

Fig. \ref{fig:RateComparisonGM} (b) shows the ARE of the simple
quadratic detector to the optimal detector as a function of
correlation strength $a$ for several values of SNR (0, 10, 20, 30
dB).  It is seen that at weak correlation the simple quadratic
detector performs as well as the optimal detector for all the
values of SNR. It is also seen that  the ARE decreases to zero
eventually as the correlation becomes strong ($a\rightarrow 1$).
This is because for the  perfectly correlated signal ($a=1$) the
optimal test statistic is in form of $(\sum_{i=1}^{n} y_i)^2$
which uses  the perfect signal correlation and adds the signal
component coherently  before taking the magnitude by squaring
\cite{Sung&Tong&Poor:05ISIT}. On the other hand, the test
statistic (\ref{eq:teststatisticsq}) for the simple quadratic
detector neglects this correlation entirely.  It is seen that the
range of correlation values over which the simple quadratic
detector performs as well as the optimal detector increases as SNR
increases. Note that at an SNR of 30 dB the simple quadratic
detector performs as well as the optimal detector through almost
the whole range of correlation except the very highly correlated
case ($ 0.9 < a \le 1$).  The behavior of ARE as a function of SNR
is summarized in the following proposition.

\vspace{0.5em}
\begin{proposition}  \label{prop:AREvsSNR}
The ARE of the simple quadratic detector
  (\ref{eq:simplequaddetector}) to the
optimal detector (\ref{eq:optimaldetector})
 increases to unity for any bounded spectrum  $f_s(\omega)$ as
SNR increases without bound.
\end{proposition}
\vspace{0.5em}

\noindent {\em Proof:}  Since the error exponent for the simple
quadratic detector is determined by the miss probability and the
optimal detector has the same error exponent for the false alarm
and miss probabilities, this can be shown via the cumulant
generating functions (\ref{eq:Lam1SQt}, \ref{eq:Lam1OPTt}) for the
two detectors. For any bounded spectrum we have
\begin{equation}
f_s(\omega) \le M, ~~\forall~0 \le \omega \le 2\pi,
\end{equation}
for some $M>0$.
 So,  we have for the second term in (\ref{eq:Lam1SQt}), as $\theta^2 \rightarrow
\infty$,
\begin{equation}
\frac{\theta^2(\sigma^2+\theta^2f_s(\omega))}{\sigma^2(\sigma^2+\theta^2)}
\rightarrow \frac{\theta^2f_s(\omega)}{\sigma^2},
\end{equation}
which is the corresponding term in (\ref{eq:Lam1OPTt}). For the
first terms in (\ref{eq:Lam1SQt}) and  (\ref{eq:Lam1OPTt}) we have
\begin{equation}
\log \frac{\sigma^2}{\sigma^2+\theta^2} \rightarrow \log
\frac{\sigma^2}{\theta^2}
\end{equation}
\begin{equation}
\int_0^{2\pi}
\log\frac{\sigma^2}{\sigma^2+\theta^2f_s(\omega)}d\omega
\rightarrow \int_0^{2\pi}
\log\frac{\sigma^2}{\theta^2f_s(\omega)}d\omega=\log
\frac{\sigma^2}{\theta^2},
\end{equation}
since $ \int_0^{2\pi} \log f_s(\omega)d\omega=0$ because of the
para-Hermitian conjugacy of the spectral factorization of
$f_s(\omega)= L(z)L^*(\frac{1}{z^*})|_{z=e^{j\omega}}$. Thus, the
two rate functions for the simple quadratic and the optimal
detectors converge as $\theta^2\rightarrow\infty$.
\hfill{$\blacksquare$}

For the spectrum (\ref{eq:spectrumGM}) we have bounded spectrum
for any fixed value of $a ~(0\le a < 1)$, which explains the
behavior of the ARE in Fig. \ref{fig:RateComparisonGM} (b) as SNR
increases.

\subsection{Triangularly Correlated Signal}

Next we consider the stationary signal with triangular
correlation, i.e.,
\begin{equation}
\Ebb\{S_0 S_k\} = \left\{ \begin{array}{ll} 1-|k|/M, &|k| < M\\
0, & |k| \ge M
\end{array} \right.
\end{equation}
where $M >0$ is the correlation length of the signal.  The
spectrum of the signal is given by the $M$th Fej\'ek kernel
\cite{Poor&Chang:85JASA}:
\begin{equation}
f_s(\omega) = \frac{1}{M} \left(
\frac{\sin(M\omega/2)}{\sin(\omega/2)} \right)^2, ~~0 \le \omega
\le 2\pi.
\end{equation}

\begin{figure}[htbp]
\centerline{ \SetLabels
\L(0.24*-0.08) (a) \\
\L(0.775*-0.08) (b) \\
\endSetLabels
\leavevmode 
\strut\AffixLabels{
\scalefig{0.4}\epsfbox{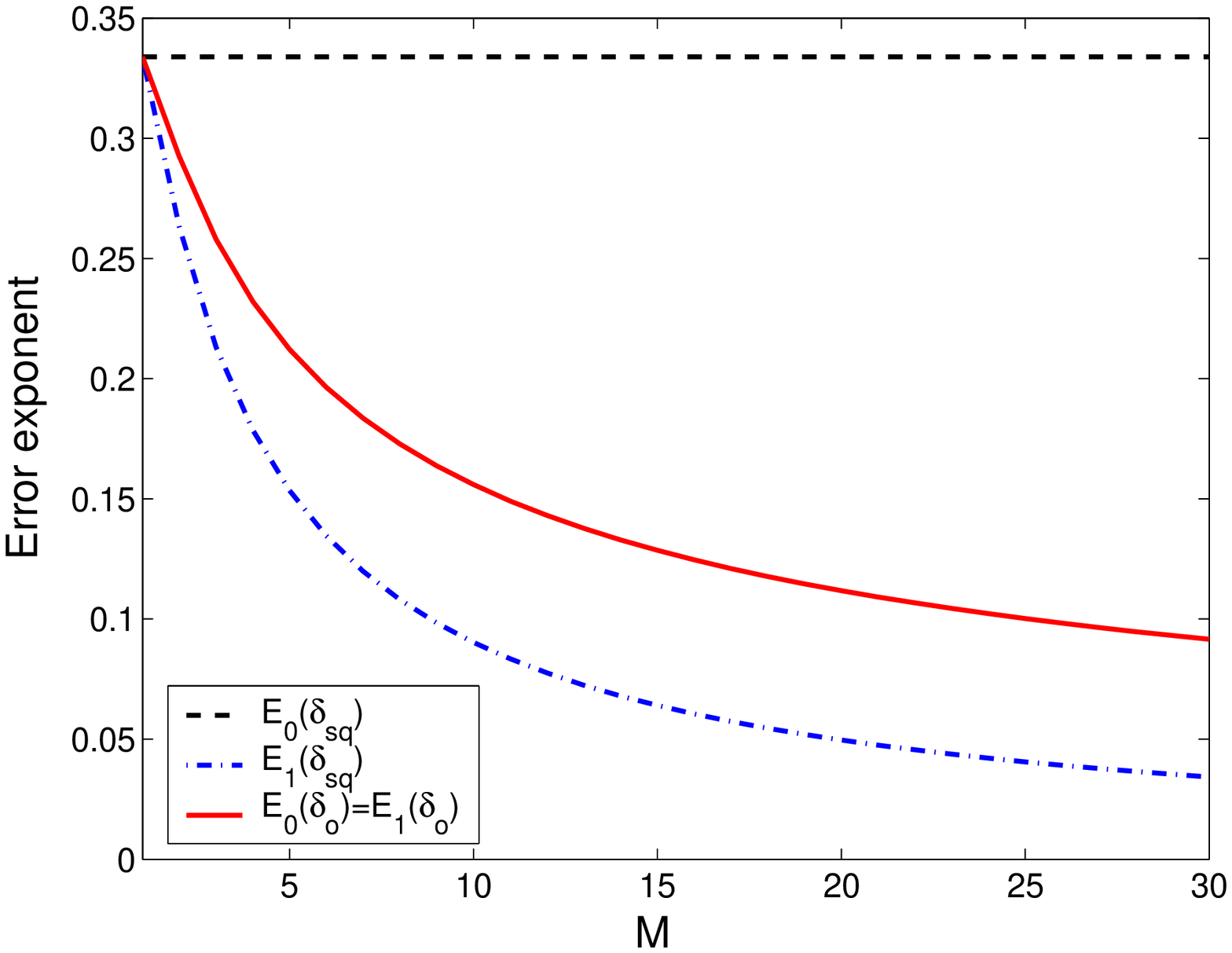}
\hspace{1cm}
\scalefig{0.405}\epsfbox{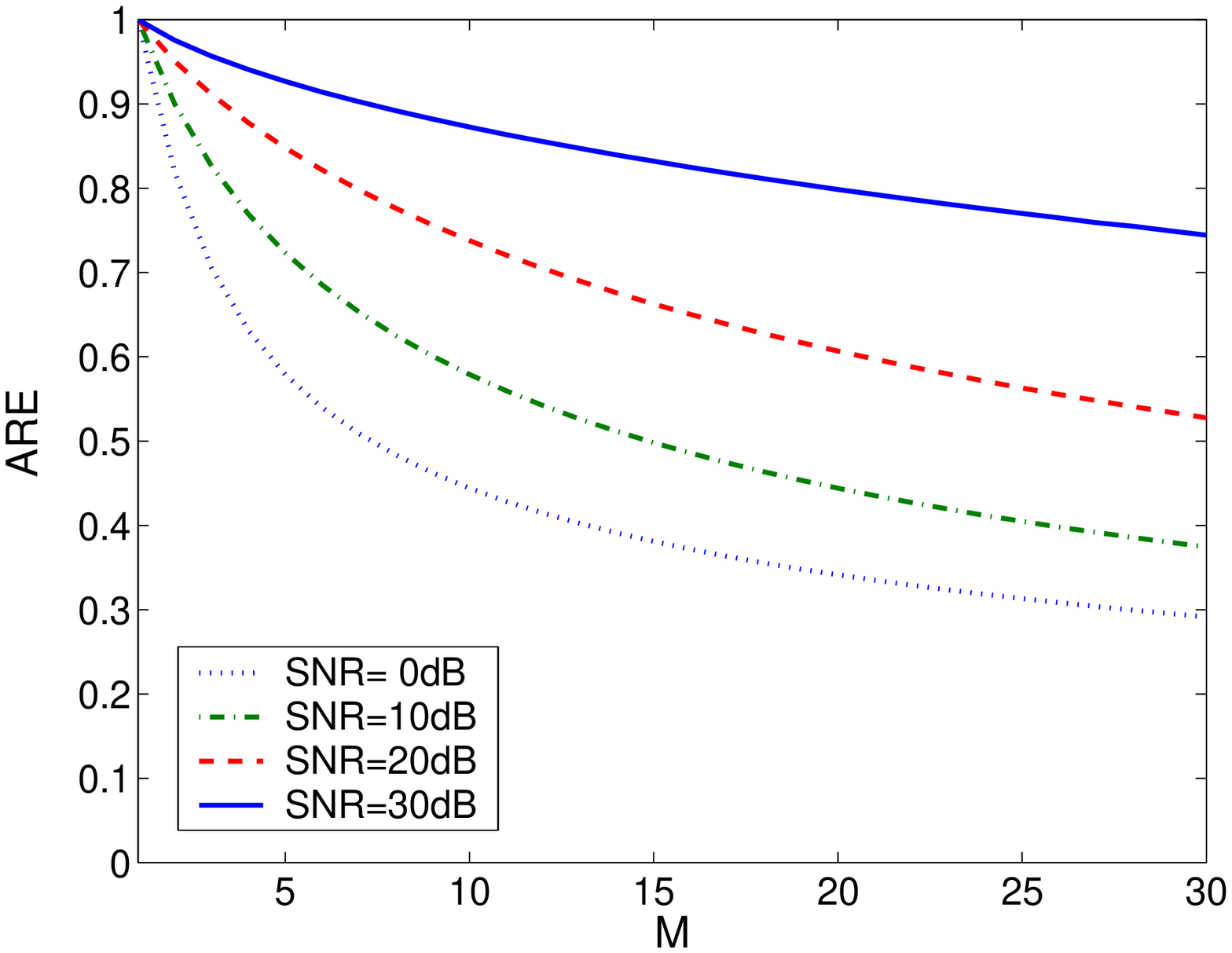} } }
 \vspace{0.1cm} \caption{Optimal and simple quadratic detectors (triangularly correlated signal): (a) error exponent,
$E_j(\delta)$, $j=0,1$, as a function of correlation strength $a$
(SNR=10dB) and (b) ARE as a function of correlation strength $a$
for SNR = 0, 10, 20, 30 dB.} \label{fig:RateComparisonTRI}
\end{figure}

Fig. \ref{fig:RateComparisonTRI} (a)  shows the error exponent for
the false alarm and miss probabilities for the optimal and simple
quadratic detectors as a function of the correlation width $M$ at
10 dB SNR for the triangularly correlated signal. Similar relative
behavior to that in the Gauss-Markov signal case is observed.  It
is worth noticing that the error exponents for the two detectors
decay sharply near $M=1$ as the correlation length $M$ increases,
and the decay is mild as $M$ further increases.

Fig. \ref{fig:RateComparisonTRI} (b) shows the ARE of the simple
quadratic detector to the optimal detector as a function of
correlation strength $a$ for the same values of SNR as in the
Gauss-Markov case.  It is seen that the ARE increases as SNR
increases as expected from Proposition  \ref{prop:AREvsSNR}.
However, at an SNR of 30 dB there exists noticeable performance
degradation for the simple quadratic detector compared with Fig.
\ref{fig:RateComparisonGM} (b) for a wide range of the correlation
length $M$.

\subsection{Banded Quadratic Detector}

\begin{figure}[hbtp]
\centerline{
\begin{psfrags}
\psfrag{Yes}[l]{Yes} \psfrag{No}[c]{No}
 \psfrag{Read}[c]{Read
$(b_0,\cdots,b_m)$} \psfrag{Compute1}[c]{Compute (50) and (51)}
\psfrag{Check1}[c]{Check (52)} \psfrag{Compute2}[c]{Compute
$I_0^{\delta_{b,m}}(0)$ and $I_1^{\delta_{b,m}}(0)$}
\psfrag{Compute3}[c]{by optimization of (53,54)}
\psfrag{Compute4}[c]{using (40,41,55,56)}
\psfrag{Compute5}[c]{Compute $t_0^*$ and $t_1^*$using ,53,54),
$t_0^*$ and $t_1^*$} \psfrag{E}[c]{$E=\min\{
I_0^{\delta_{b,m}}(0),I_1^{\delta_{b,m}}(0)\}$}
\psfrag{Store}[c]{Store $(b_0,\cdots,b_m, E)$}
\scalefig{0.4}\epsfbox{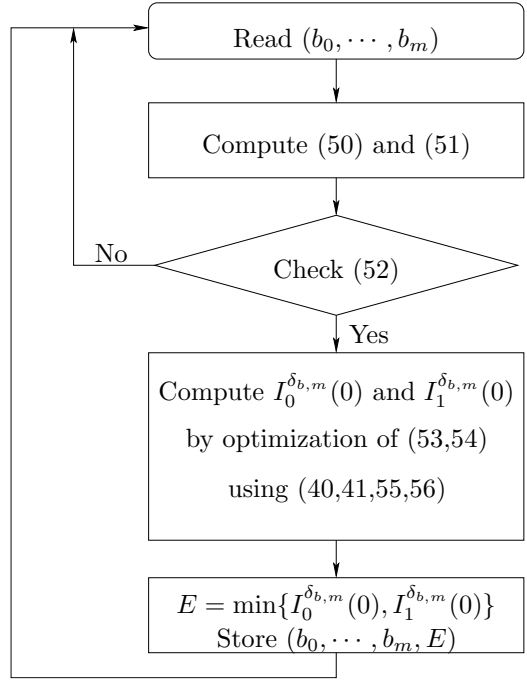}
\end{psfrags}
} \caption{An optimization algorithm (grid search).}
\label{fig:algorithm}
\end{figure}

We here provide some necessary conditions for the optimal
$\tilde{\Qbf}_n^{(m)}$ in (\ref{eq:bandedquadLn}) and evaluate the
performance of the banded quadratic detector.  The test statistic
(\ref{eq:bandedteststatsumform}) has two different limits (as
$n\rightarrow \infty$) under the two hypotheses, and they are
given by
\begin{eqnarray}
\bar{T}_0^{(m)} &\defeq& \lim_{n\rightarrow\infty}
\{T_{b,n}^{(m)}|H_0\}= \frac{1}{4\pi}\int_0^{2\pi}
\log\frac{\sigma^2}{\sigma^2+\theta^2f_s(\omega)}d\omega+  \frac{1}{4\pi}\int_0^{2\pi}\sigma^2 g_m(\omega) d\omega,\label{eq:bandedlimT0}\\
\bar{T}_1^{(m)} &\defeq& \lim_{n\rightarrow\infty}
\{T_{b,n}^{(m)}|H_1\}= \frac{1}{4\pi}\int_0^{2\pi}
\log\frac{\sigma^2}{\sigma^2+\theta^2f_s(\omega)}d\omega+
\frac{1}{4\pi}\int_0^{2\pi}(\sigma^2+\theta^2f_s(\omega))
g_m(\omega) d\omega.\label{eq:bandedlimT1}
\end{eqnarray}
The first term in each equation is by applying Theorem
\ref{theo:ToeplitzDT}, and the second term follows from the law of
large numbers and  $\frac{1}{n}\Ebb_j \{ \ybf_n^T
\tilde{\Qbf}_n^{(m)}\ybf_n\} = \frac{1}{n}
\mbox{tr}\{\tilde{\Qbf}_n^{(m)} \Sigmabf_{j,n}\}=\frac{1}{n}
\sum_{i=1}^n \tilde{\lambda}_{ij}^{(n)}$ (to which
(\ref{eq:ToeplitzProduct}) is applied) since the trace of a matrix
is the sum of its eigenvalues.  From (\ref{eq:bandedlimT0},
\ref{eq:bandedlimT1}) we have $\bar{T}_0^{(m)} < \bar{T}_1^{(m)}$
for any $\theta >0$ and a signal spectrum which is not identically
zero.  One necessary condition for the optimal $g_m(\omega)$ is
given by
\begin{equation}  \label{eq:bandedtQcondition}
\bar{T}_0^{(m)} < \tau ~(=0) < \bar{T}_1^{(m)}.
\end{equation}
Otherwise, the error exponent $E(\delta_{b,m})$ is zero and the
average error probability of the banded-quadratic detector decays
 at subexponential rate as $n$ increases. For example, if
$\bar{T}_0^{(m)}
> 0$, then $E_0(\delta_{b,m})=\inf_{z \in [0,\infty)}$
$I_0^{\delta_{b,m}}(z) =0$ since
$I_0^{\delta_{b,m}}(\bar{T}_0^{(m)})=0$. Similarly, we have
$E_1(\delta_{b,m})=\inf_{z \in (-\infty,0)} I_1^{\delta_{b,m}}(z)
=0$ if $\bar{T}_1^{(m)} < 0$.  Thus, in the case of $m=0$ we have
$g_m(\omega)=b_0$ and it is seen from (\ref{eq:bandedlimT1}) that
the optimal $b_0$ is positive (otherwise, $\bar{T}_1^{(0)} < 0$),
which is consistent with our assumption of the
positive-definiteness of $\tilde{\Qbf}_n^{(m)}$.
 In general, it is easy to see that
well chosen $b_0,\cdots,b_m$ satisfy the condition
(\ref{eq:bandedtQcondition}) since the first terms in
(\ref{eq:bandedlimT0}, \ref{eq:bandedlimT1}) are equivalent and
negative.  When (\ref{eq:bandedtQcondition}) is satisfied, it is
known that the infimum for the rate function is achieved at the
decision threshold,\cite{Dembo&Zeitouni:book} i.e.,
\begin{eqnarray}
E_0(\delta_{b,m})&=&\inf_{z \in [0,\infty)} I_0^{\delta_{b,m}}(z)=
I_0^{\delta_{b,m}}(0)= \sup_{-\infty < t < \inf_{0\le \omega \le
2\pi} \left\{ \frac{1}{\sigma^2g_m(\omega)}\right\}}
\{-\Lambda_0^{\delta_{b,m}}(t) \},
\label{eq:E0deltabm}\\
E_1(\delta_{b,m})&=&\inf_{z \in (-\infty,0)}
I_1^{\delta_{b,m}}(z)=I_1^{\delta_{b,m}}(0)=\sup_{-\infty < t <
\inf_{0\le \omega \le 2\pi} \left\{ \frac{1}{
(\sigma^2+\theta^2f_s(\omega))g_m(\omega)}\right\}}
\{-\Lambda_1^{\delta_{b,m}}(t) \},\label{eq:E1deltabm}
\end{eqnarray}
where $\Lambda_0^{\delta_{b,m}}(t)$ and
$\Lambda_1^{\delta_{b,m}}(t)$ are given by (\ref{eq:bandedLam0})
and  (\ref{eq:bandedLam1}), respectively, and the optimal values
of $t$ for (\ref{eq:E0deltabm}) and (\ref{eq:E1deltabm}) are given
by solving
\begin{equation}
\frac{1}{4\pi}\int_0^{2\pi}
\log\frac{\sigma^2}{\sigma^2+\theta^2f_s(\omega)}d\omega+\frac{1}{4\pi}\int_0^{2\pi}\frac{\sigma^2g_m(\omega)}{
1-t_0^* \sigma^2 g_m(\omega)}d\omega=0
\end{equation}
and
\begin{equation} \label{eq:bandedopttstar1}
\frac{1}{4\pi}\int_0^{2\pi}
\log\frac{\sigma^2}{\sigma^2+\theta^2f_s(\omega)}d\omega+\frac{1}{4\pi}\int_0^{2\pi}\frac{(\sigma^2+\theta^2f_s(\omega))
g_m(\omega)}{ 1-t_1^* (\sigma^2+\theta^2f_s(\omega))
g_m(\omega)}d\omega=0,
\end{equation}
 respectively.
  Thus, the optimal
$g_m(\omega)$ for given $m$, SNR and signal spectrum is obtained
from the following optimization problem:
\begin{equation} \label{eq:bandedoptimization}
g_m^*(\omega) = \mathop{\arg \max}_{b_0,\cdots,b_m} \left\{ \min
\{I_0^{\delta_{b,m}}(0), I_1^{\delta_{b,m}}(0) \}\right\}
\end{equation}
under the constraint (\ref{eq:bandedtQcondition}).  A closed-form
expression for (\ref{eq:bandedoptimization}) seems difficult to
obtain in general cases.  However,
(\ref{eq:bandedtQcondition}-\ref{eq:bandedopttstar1}) facilitate
numerical approaches to the optimization problem, and a procedure
using grid search is summarized in Fig. \ref{fig:algorithm}.

\begin{figure}[hbtp]
\centerline{
\begin{psfrags}
\psfrag{Sensori}[c]{\textsf{Sensor} $i$}
\psfrag{Cim1}[l]{$C_{i-1}$} \psfrag{yim1}[l]{$y_{i-1}$}
\psfrag{Ci}[l]{$C_{i}$} \psfrag{yi}[l]{$y_{i}$}
\psfrag{Ci=}[c]{$C_{i}=C_{i-1}+2b_1y_{i-1}y_i + b_0y_i^2$}
\scalefig{0.5}\epsfbox{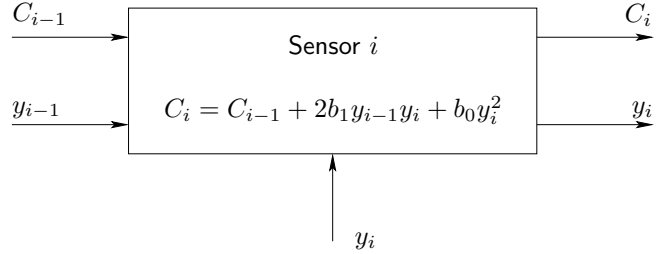}
\end{psfrags}
} \caption{Banded-quadratic detector with $m=1$: Distributed
computation in wireless sensor network ($C_0=b_0y_1^2$).}
\label{fig:bandedquadm1}
\end{figure}

\begin{figure}[htbp]
\centerline{ \SetLabels
\L(0.24*-0.08) (a) \\
\L(0.775*-0.08) (b) \\
\endSetLabels
\leavevmode 
\strut\AffixLabels{
\scalefig{0.407}\epsfbox{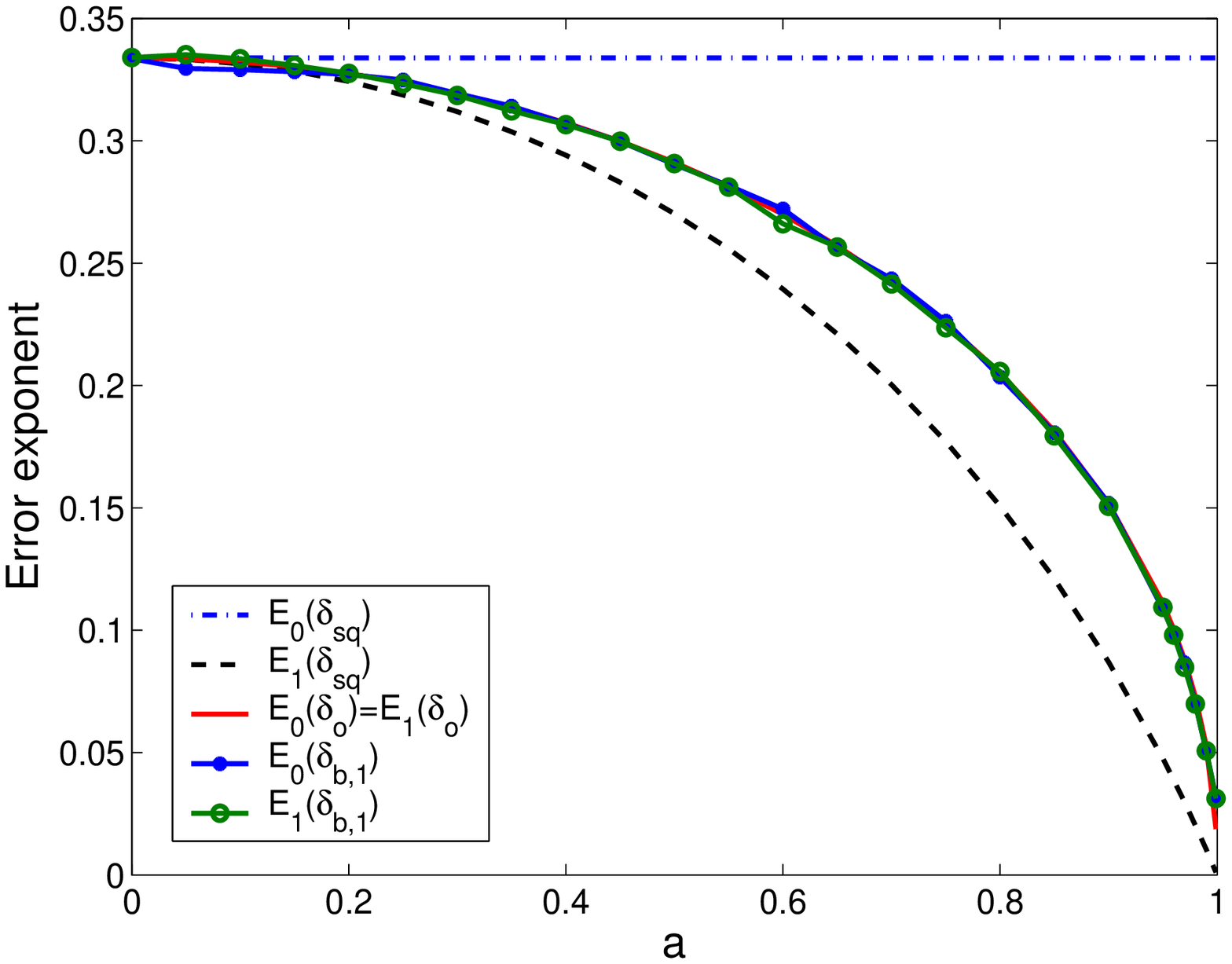}
\hspace{1cm}
\scalefig{0.4}\epsfbox{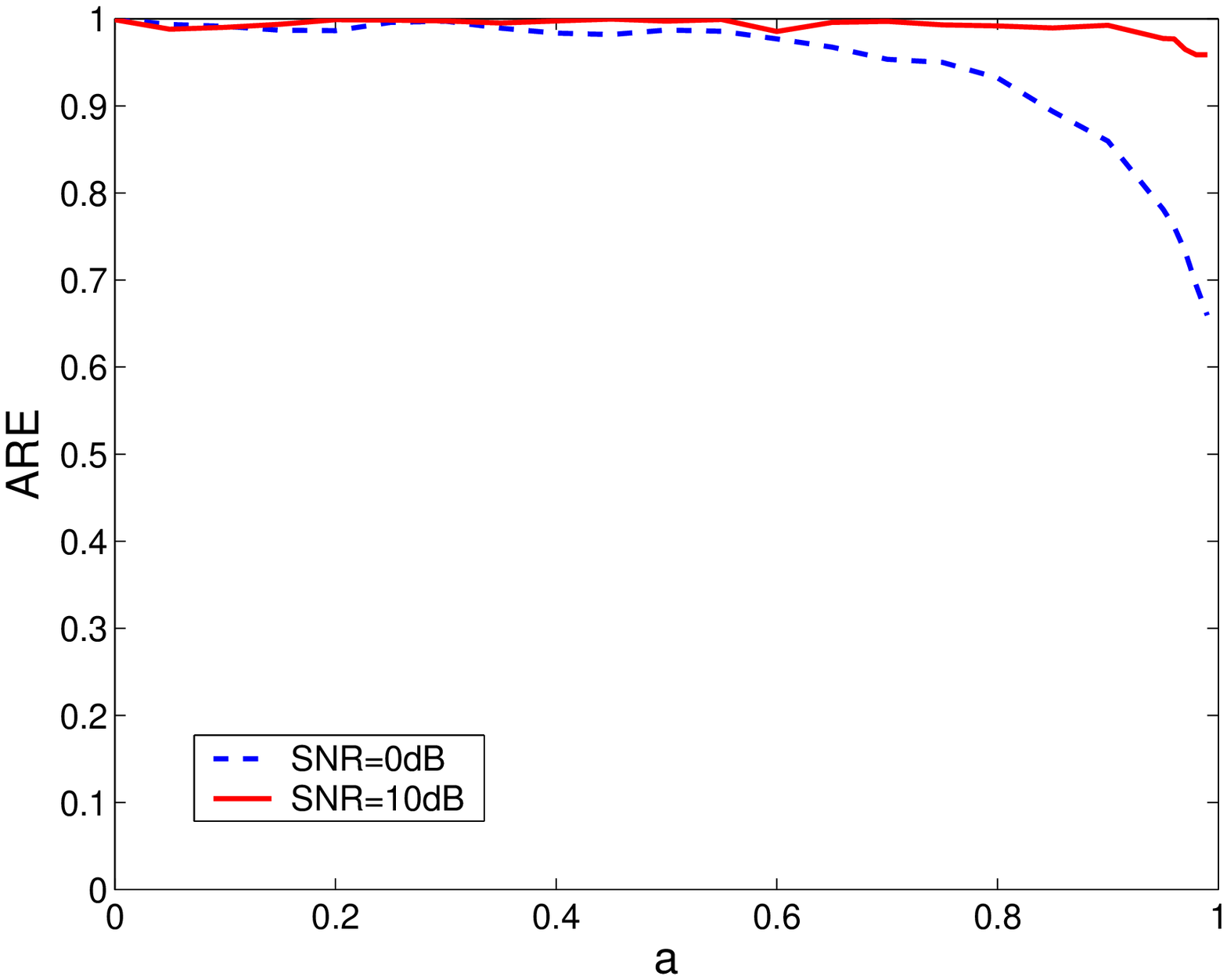}}}
 \vspace{0.1cm} \caption{Banded-quadratic detector (Gauss-Markov signal): (a) error exponent,
$E_j(\delta)$, $j=0,1$, as a function of correlation strength $a$
for SNR = 10 dB and (b) ARE as a function of correlation strength
$a$ with the optimized $g_m(\omega)$ (SNR = 0, 10 dB).}
\label{fig:RateComparisonGMBanded}
\end{figure}

We considered the Gauss-Markov signal (\ref{eq:spectrumGM}) and
evaluated the banded-quadratic detector with $m=1$ which
corresponds to the case that each sensor requires the information
only from a neighboring sensor in a wireless sensor network setup,
as shown in Fig. \ref{fig:bandedquadm1}. Fig.
\ref{fig:RateComparisonGMBanded} (a) shows the error exponents
$E_0(\delta_{b,1})$ and $E_1(\delta_{b,1})$ of the
banded-quadratic detector optimized using the algorithm shown in
Fig. \ref{fig:algorithm} for each value of $a$ at 10 dB SNR. Fig.
\ref{fig:RateComparisonGMBanded} (b) shows the corresponding ARE
of the banded-quadratic detector to the optimal detector. For a
SNR of 0 dB SNR the banded-quadratic detector with $m=1$ performs
well not only in the low correlation values but also in the high
correlation region where the performance of the simple quadratic
detector degrades severely (see Fig. \ref{fig:RateComparisonGM}
(b)). Surprisingly, it is seen that optimal performance is almost
achieved with only $m=1$ for a SNR of 10 dB.

\section{Conclusions}
\label{sec:conclusion}

 We have considered the relative performance of several quadratic detectors for
Gaussian signals in Gaussian noise under a Bayesian formulation.
Using the large deviations principle, a general form of the rate
function for the simple quadratic detector, optimal detector, and
banded-quadratic detector has been provided using the signal
spectrum. For the examples of Gauss-Markov and triangularly
correlated signals we have evaluated the error exponents for the
false alarm and miss probabilities and the ARE for the average
error probability. We have also investigated the effects of SNR on
the relative performance. The asymptotic efficiency of the simple
quadratic detector relative to the optimal detector converges to
unity as SNR increases without bound for any bounded signal
spectrum. At high SNR the simple quadratic detector performs as
well as the optimal detector for a wide range of correlation
values and the banded-quadratic detector effectively achieves the
optimal performance with much lower complexity.

\acknowledgments     

This work was supported in part by the Multidisciplinary
University Research Initiative (MURI)  under the Office of Naval
Research Contract N00014-00-1-0564.  Prepared through
collaborative participation in the Communications and Networks
Consortium sponsored by the U.~S. Army Research Laboratory under
the Collaborative Technology Alliance Program, Cooperative
Agreement DAAD19-01-2-0011.

The work of H. V. Poor was supported in part by the Office of
Naval Research under Grant N00014-03-1-0102.


\end{document}